\documentclass[11pt]{entcs/entcs} \usepackage{entcs/entcsmacro}

\usepackage{graphicx}
\usepackage{mathtools} 
\usepackage{enumerate}
\usepackage{stmaryrd}
\usepackage{amsmath}
\usepackage{caption}
\usepackage{subcaption}
\usepackage[title]{appendix}

\def\lastname{Beica, Feret and Petrov}
\begin{document}

\begin{frontmatter}
  \title{Tropical Abstraction of Biochemical Reaction Networks with Guarantees} 
  
  \author{Andreea Beica
          \thanksref{myemail}}
  \address{D\'{e}partement d'informatique  \\ \'{E}cole normale sup\'{e}rieure, CNRS, PSL Research University \\75005 Paris, France} 
  
  \author{J\'{e}r\^{o}me Feret\thanksref{coemail}} \address{D\'{e}partement d'informatique  \\ \'{E}cole normale sup\'{e}rieure, CNRS, PSL Research University \\75005 Paris, France \\
INRIA} 

  \author{Tatjana Petrov \thanksref{co2email}} \address{Department of Computer and Information Science \\ University of Konstanz, Germany}

    \thanks[myemail]{Email:
    \href{mailto:beica@di.ens.fr} {\texttt{\normalshape
        beica@di.ens.fr}}} 
        
    \thanks[coemail]{Email:
    \href{mailto:feret@di.ens.fr} {\texttt{\normalshape
        feret@di.ens.fr}}}
        
    \thanks[co2email]{Email:
    \href{mailto:tatjana.petrov@uni-konstanz.de} {\texttt{\normalshape
        tatjana.petrov@uni-konstanz.de}}, Tatjana Petrov's research is supported by  the Ministry of Science, Research and the Arts of the state of Baden-Württemberg}
\begin{abstract}
Biochemical molecules interact through modification and binding
reactions, giving raise to a combinatorial number of possible biochemical
species. The time-dependent evolution of concentrations of the species is
commonly described by a system of coupled ordinary differential equations
(ODEs). However, the analysis of such high-dimensional, non-linear system of 
equations is often computationally expensive and even prohibitive in 
practice. The major challenge towards reducing such models is providing the
guarantees as to how the solution of the reduced model relates to that of
the original model, while avoiding to solve the original model.

In this paper, we have designed and tested an approximation method for ODE models of biochemical reaction systems, in which the guarantees are our major requirement. Borrowing from tropical analysis techniques, we look at the dominance relations among terms of each species' ODE. These dominance relations can be exploited to simplify the original model, by neglecting the dominated terms. As the dominant subsystems can change during the system's dynamics, depending on which species dominate the others, several possible modes exist. Thus, simpler models consisting of only the dominant subsystems can be assembled into hybrid, piecewise smooth models, which approximate the behavior of the initial system. 
By combining the detection of dominated terms with symbolic bounds propagation, we show how to approximate the original model by an assembly of simpler models, consisting in ordinary differential equations that provide time-dependent lower and upper bounds for the concentrations of the initial model’s species.

The utility of our method is twofold. On the one hand, it provides a reduction heuristics that performs without any prior knowledge of the initial system's behavior (\emph{i.e.}, no simulation of the initial system is needed in order to reduce it). 
On the other hand, our method provides sound interval bounds for each species, and hence can serve to evaluate the faithfulness of tropicalization reduction heuristics for ODE models of biochemical reduction systems. 
The method is tested on several case studies.


\end{abstract}
\end{frontmatter}

\section{Introduction}

As biology becomes a data intensive science, due to advancements in high throughput molecular biology, the importance of \textit{in silico} dynamical models of complex biological systems that are able to reproduce intricate behaviors observed in experimental settings increases. As such, modeling becomes a part of biological reasoning, but turns out to be a particularly challenging task. 
In particular, in models of biochemical networks, the number of possible chemical species is often subject to combinatorial explosion, due to the large 
number of species that may arise as a result of protein bindings and post-translational modifications \cite{hlavacek2003complexity}. 
As a consequence, mechanistic models of signaling pathways easily become very combinatorial.
%
%
A common modeling approach describes the time-dependent evolution of concentrations of each of the modeled species through a system of coupled ordinary differential equations (ODEs).
The combinatorial explosion of species and rich interaction scheme renders solving such a system of ODEs often prohibitory in practice, let alone the fact that is it already an approximation of its stochastic counter-part \cite{kurtz1970solutions}, as well as that the equations themselves do not transparently reflect the underlying mechanisms.
Addressing the latter, formalisms allowing to write the mechanistic hypothesis in form of discrete transition steps have been proposed:
Boolean networks\cite{wang}, logical networks\cite{wyn}, Petri Nets\cite{chaouiya}, cellular automata\cite{green}, rule-based languages\cite{kappa1}, to name the most common. Languages such as Kappa\cite{kappa1,kappa2} and BNGL\cite{bngl} provide compact ways of describing models prone to combinatorial explosion, of simulating them \cite{kappa3}, and even transforming them into ODEs \cite{bngl}.
However, the curse of dimensionality once again rises when trying to compute the system behavior.
 


A strategy to cope with such complexity is model reduction, in which certain properties of biochemical models are exploited in order to obtain simpler versions of the original complex model; these simpler models should preserve the important behavioral aspects of the initial system.
An example of such a property is the multiscaleness of biochemical networks, with respect to both time-scales and species' abundance. In the case of the former, it is known that biochemical processes governing network dynamics span over many well separated timescales: while protein complex formation occurs on the seconds scale, post-translational protein modification takes minutes, and changing gene expression can take hours, or even days. As for the latter, multiscaleness also applies to the abundance of various species in biochemical networks: the DNA molecule has one to a few copies, while  mRNA copy numbers can vary from a few to  tens of thousands. 
On the one hand, these widely different time- and concentration scales represent challenges for the estimation of rate constants, for the measurement of low-concentration species, and even for the numerical integration. 
On the other hand, they represent a feature that can be exploited for model reduction purposes, allowing to approximate the complete mechanistic description with simpler rate expressions, retaining the essential features of the full problem on the time scale or in the concentration range of interest. 
The dynamics of multiscale, large biochemical systems can be reduced to those of simpler models, called \emph{dominant} subsystems \cite{vnoel}, which contain less parameters and are easier to analyze. Dominant subsystems are chosen by comparing the time-scales of the large system.  For example, the classical quasi steady-state (QSS) \cite{qss} and quasi-equilibrium (QE) approximations \cite{mm,qe} are conditions that lead to dominance, and represent popular methods for the computation of ``first approximations'' to the slow invariant manifold. 
Classical QSS is based on the small concentrations of highly reactive intermediate species (\textit{i.e.}, atoms, ions, enzymes and substrate-enzyme complexes)\cite{boseung}, while in the QE approximation the reduction of the full mechanism is done based on the existence of \textit{fast} and \textit{slow} reactions.  

The multiscaleness property of biochemical network is by definition closely linked to the mathematical notion of dominance, captured in the framework of tropical analysis\cite{max-plus,lit}. 
Recently, a class of semi-formal methods for reducing and hybridizing models of biochemical networks has been developed, based on ideas from tropical analysis \cite{vnoel,vnoel2,vnoel3,rad}. These methods exploit the multiscaleness of biochemical networks, in order to deduce \textit{dominance} relations among parameters and/or reaction rates, which can then be used to obtain a system of truncated ODEs (by eliminating the dominated terms). One of the advantages of using dominance relations in multi-scale networks is that it helps cope with parameter uncertainty: parameter values are replaced with their orders of magnitude, which are easier to determine. 
However, providing guarantees as to how the solution of the reduced model relates to the original one remains a challenge. Such is also the case in \cite{beica}, where we proposed a reduction framework for rule-based models, based on time-scale separation. While this time-scale separation technique is justified by asymptotic convergence results, for any concrete parameter values, there is no information on the accuracy of the trajectories obtained by executing the reduced model.

In this paper, we design an approximation method for ODE models of biochemical networks, in which the guarantees are our major requirement. Our method combines abstraction and numerical approximation, and aims at providing better understanding/evaluation of tropical reduction methods. We abstract the solution of the original system of ODEs by a box, that over-approximates the state of the original system and provides lower and upper bounds for the value of each variable of the system in its current state. The simpler equations (which we call \emph{tropicalized}) defining the hyperfaces of the box are obtained by combining the dominance concept borrowed from tropical analysis with symbolic bounds propagation. Mass invariants of the initial system of ODEs are used to refine the computed bounds, thus improving the accuracy of the method. The resulting (simplified) system provides \emph{a posteriori} time-dependent lower and upper bounds for the concentrations of the initial model's species, and thus bounds on numerical errors stemming from tropicalization. This means that no information on the original system's trajectory is needed - the most important advantage of our approach. By contrast, the main difficulty of applying the classical QSS and QE reductions to biochemical models is that QE reactions and QSS species need to be specified \emph{a priori}, which implies that some knowledge about the initial system's behavior is necessary. This, in turn, means that significantly high-dimensional, non-linear systems cannot benefit from these reductions, as their analysis can be prohibitive in practice.
An approach similar with respect to providing \emph{a posteriori}  time-dependent lower and upper bounds has been proposed in \cite{ken}, where  the differential semantics of rule-based models with non-contracting dynamics and unbounded sets of variables are treated. Rather than using dominance relations between ODE terms, a finite set of patterns is used in order to bound the number of occurrences of each pattern. Further related works, similar in the sense that they provide automatizible reduction methods with strong reduction guarantees are described in \cite{feret2009internal,feret2012lumpability}. However, both of these works are designed specifically for rule-based models, where they exploit the site-graph encoding of species' structure, rather than the dominance regions.

 Depending on the chosen granularity of mass-invariant-derived bounds, the method presented in this paper can either be used to reduce models of biochemical networks, or to quantify the approximation error of tropicalization reduction methods that do not involve guarantees.
The guarantees of our method are obtained by formalizing the soundness relation between the original system of equations and the abstract system of ordinary differential equations operating on the coordinates of the hyper-faces of the box. The solution of a sound abstraction of an original system of differential equations, starting from a box that contains the initial state of the original system, defines a sound abstraction of the solution(s) of the original system. We apply our method to several case studies.

\textbf{Outline} The rest of this article is organized as follows. In Sect.2 we define the setting and concepts used in our approach, as well as introduce motivating examples. We then formally present and justify the method for deriving the system of reduced ODEs over the lower and upper bounds of species' concentrations in Sect. 3. Also in Sect. 3, we present the two possible uses of our approach. We then discuss and conclude in Sect. 4.

\section{Definitions and Motivating Example(s)}\label{sect:2}

\subsection{General Setting and Definitions}

We define a dynamic reaction network over a set of species $\mathcal{S}=\{x_1,\ldots,x_s\}$ as a reaction system of the form:

\begin{equation}
r_j: \sum\limits_{i}\alpha_{ji}x_i \xleftrightarrow{k_j^+,k_j^-} \sum\limits_{i} \beta_{ji} x_i, 
\end{equation}
where $1\leq j \leq r$ is the reaction number, and for each reaction $r_j$, $k_j^{+/-}$ are the non-negative reaction rate constants of the forward, respectively backward reaction.

In this paper, we focus on the ordinary differential equations (ODE) semantics of models of biochemical networks. The underlying assumptions are that the various species of the chemical network are highly abundant, that stochastic fluctuations are negligible, and that the reaction system is well-mixed. In these conditions, the state of the dynamic system (1)  can be represented as a multiset of the species' concentrations $\textbf{x} \in \mathbb{R}^n$, and its dynamics is described by a system of ODEs:

\begin{equation}
\frac{d\textbf{x}}{dt}=\sum\limits_{j=1}^{r}(\mathbb{\boldsymbol{\beta}}_j-\mathbb{\boldsymbol{\alpha}}_j)(R_j^+(\textbf{x})-R_j^-(\textbf{x})).
\end{equation}

For each reaction $r_j$, $R^{+/-}_j(\textbf{x})$ denotes the reaction (forward and backward) rate, and is a non-linear function of the concentrations. For example, the mass-action law reads:

\begin{equation}
\begin{split}
& R^+_j(\textbf{x})=k_j^+\prod_i x_i^{\alpha_{ji}} \\
& R^-_j(\textbf{x})=k_j^-\prod_i x_i^{\beta_{ji}},
\end{split}
\end{equation}
which in turn means that $\frac{d\textbf{x}}{dt}$ is a multivariate polynomial of the species' concentrations. In other words, the ODE of the $i$-th species, $x_i$, under mass action kinetics reads as a sum of monomials, which can be split into \emph{production} and \emph{consumption} terms, according to the sign that precedes their occurence in the equation:

\begin{equation}\label{eq:ODE}
\frac{dx_i}{dt} = P_i^+(\textbf{x}) - P_i^-(\textbf{x}),
\end{equation}

where $P_i^{+/-}(\textbf{x})$ are Laurent polynomials with positive coefficients: 

\begin{equation}
\begin{split}
&P_i^+(\textbf{x}) = 
\sum\limits_{\beta_{ji}-\alpha_{ji}> 0} (\beta_{ji}-\alpha_{ji})R_j^+(\textbf{x}) + 
\sum\limits_{\beta_{ji}-\alpha_{ji}<0} (\alpha_{ji}-\beta_{ji})R_j^-(\textbf{x})\\
&P_i^-(\textbf{x}) = 
\sum\limits_{\beta_{ji}-\alpha_{ji}< 0} (\alpha_{ji}-\beta_{ji})R_j^+(\textbf{x}) + 
\sum\limits_{\beta_{ji}-\alpha_{ji}>0}(\beta_{ji}-\alpha_{ji})R_j^-(\textbf{x})
\end{split}
\end{equation}

For convenience purposes, we will denote $P_i^{+/-}(\textbf{x}) = \sum\limits_j M_{i,j}^{+/-}(\textbf{x})$, where $M_{i,j}^{+/-}(\textbf{x})$ represent the production, respectively the consumption, monomials.

The reduction heuristics that use ideas from tropical analysis exploit the concept of dominance, which we borrow for our method. Let $M_1(\textbf{x})=c_1\textbf{x}^{\alpha_1}$ and  $M_2(\textbf{x})=c_2\textbf{x}^{\alpha_2}$  be two (positive) monomials. We define $\epsilon$-dominance as the following partial order relation on the set of multivariate monomials defined on subsets of $\mathbb{R}_+^n$:

\begin{definition}\label{def:dom}\textbf{($\epsilon$-dominance)}
For an $\epsilon\in[0,1]$, we say that $M_1$ dominates $M_2$ at a time point $t$, denoted by $M_1\succ_\epsilon M_2$, if $\epsilon \cdot  M_1(\textbf{x}(t)) \geq M_2(\textbf{x}(t))$.
\end{definition}

In multiscale biochemical systems, the various monomials that compose the polynomials $P_i^{+/-}$ have different magnitude orders, such that at any given time there is only one or a few dominating monomials.

\begin{definition}\textbf{(Dominant monomial of a polynomial)}
For a given $\epsilon\in[0,1]$, the dominant monomial of a polynomial $P_i(\textbf{x})=\sum\limits_{j=1}^n M_{i,j}(\textbf{x})$ is defined as $Dom(P_i)=\{M_{i,j} \mid \forall 1\leq k\leq n, j\neq k, M_{i,j}\succ_\epsilon M_{i,k}\}$.
\end{definition}

  By using the max-plus algebra idea that the sum of positive, well separated terms, can be replaced by the maximum term, each of the two polynomials of (\ref{eq:ODE}) can be replaced by their dominant monomials. The result is a reduced model, consisting of a piecewise smooth function. As the dominant monomials of the $P_i^{+/-}$ can change from one concentration domain to another, the reduced model is a piecewise-smooth \emph{hybrid} model.

\begin{definition}\textbf{(Two-term tropicalization of the smooth ODE system)}
We call two-term tropicalization of the smooth ODE system (4) the following piecewise-smooth system:

\begin{equation}
\frac{dx_i}{dt}=Dom(P_i^+(\textbf{x})) - Dom(P_i^-(\textbf{x}))
\end{equation}
\end{definition}

We note that a \emph{one-term tropicalization} of the smooth ODE system, $Dom(\frac{dx_i}{dt})$, is also possible, but choosing only one dominant momomial instead of the production-consumption pair of dominant monomials leads to a less precise model reduction (as more information is discarded in the one-term tropicalization). Thus, in this paper, we choose to deal with the two-term method.

\subsection{Motivating example: Michaelis-Menten}

In Sect.1, we mentioned that the classical QSS\cite{qss} and QE\cite{qe,mm} approximations represent popular methods for the simplification of biochemical networks.  

As such, our motivating example is the Michaelis-Menten mechanism.  This enzymatic model illustrates how these two simple methods that use the idea of \emph{dominance} can be useful for model reduction of nonlinear models with multiple timescales.  The Michaelis-Menten mechanism consists of an enzyme that reversibly binds a substrate to form a complex, which in turn releases a product, while preserving the enzyme:
\begin{equation}\label{sys:orig}
\mathrm{E} + \mathrm{S} \xrightleftharpoons[k_{-1}]{k_1} \mathrm{E:S} \xrightarrow{k_2}\mathrm{E} + \mathrm{P} 
\end{equation}

The ODE system describing the evolution of the species' concentration writes as:
\begin{equation}\label{eq:orig}
\begin{cases}
\frac{d[S]}{dt} = k_{-1}[E:S] - k_{1}[E][S]\\
\frac{d[E]}{dt} = k_{-1}[E:S] + k_2[E:S] - k_{1}[E][S] \\
\frac{d[E:S]}{dt} = k_{1}[E][S] - k_{-1}[E:S] - k_2[E:S]\\
\frac{d[P]}{dt} = k_2[E:S]
\end{cases}
\end{equation}

The Michaelis-Menten equation relates the rate of product formation to the substrate concentration: 
\begin{equation}\label{eq:MM-QSS}
v=\frac{d[P]}{dt}=k_2 \frac{[E]_T [S]}{K_M+[S]},
\end{equation}
where $[E]_T=[E]+[E:S]$ is the total enzyme concentration, and $K_M=\frac{k_2+k_{-1}}{k_1}$ is the Michaelis-Menten constant. Eq.(\ref{eq:MM-QSS})can be interpreted as the reaction rate of a reduced reactive system, equivalent to system (\ref{sys:orig}), in which the intermediary complex $[E:S]$ has been eliminated:

\begin{equation}\label{sys:MM-QSS}
\mathrm{S}\xrightarrow{k_2 \frac{[E]_T}{K_M+[S]}} \mathrm{P}
\end{equation}

The approximation of (\ref{eq:orig}) by (\ref{eq:MM-QSS}) is generally considered to be sufficiently good if the QSS assumption holds, that is if the total initial enzyme concentration is much lower than the total initial concentration of substrate: $[S]_0\succ_\epsilon[E]_0+[E:S]_0$. In this case, the complex $[E:S]$ is a low concentration fast species, whose concentration is dominated by that of the substrate; the value of $[E:S]$ almost instantly relaxes to a value determined by $[S]$. Thus, one can set $\frac{d[E:S]}{dt}=0$, and exploit this relation to pool the two reactions of the initial system (\ref{sys:orig}) into an unique irreversible reaction (\ref{sys:MM-QSS}). The QSS condition can also be stated as $k_2\gg k_{-1}$ \cite{vnoel}.

The original MM analysis used the complementary QE approximation, which considers the  complex formation reaction to be \emph{fast} and \emph{reversible}: $k_{-1}\gg k_2$. Thus, the term $k_{-1}[E:S]$ dominates the term $k_2[E:S]$ in Eq.(\ref{eq:orig}), meaning the latter can be discarded from the ODE system, allowing for pooling of species, and resulting once again in a single step approximation that reads:

\begin{equation}\label{sys:MM-QE}
\mathrm{S}\xrightarrow{k_2 \frac{[E]_T}{K_d+[S]}} P,
\end{equation}

with $K_d=\frac{k_{-1}}{k_1}$, if indeed $[S]\gg[E]+[E:S]$. We note that if the QE assumption is indeed valid, $K_M\approx K_d$. 

One of the main difficulties of applying QSS and QE reductions to biochemical models is that the QE reactions and QSS species need to be specified a priori. Thus, simulation of the original model is sometimes \footnote{In \cite{vnoel2}, the authors propose a formal method for the identification of QSS species and QE conditions, which follows from the calculation of the tropicalized system, and which does not require simulation of trajectories.} needed in order to detect dominated  species, which are either QSS species, or participate in QE reactions \cite{vnoel}. For high-dimensional non-linear systems, this requirement can represent an obstacle towards model reduction. 

The issue regarding simulation of the initial system also arises when trying to quantify the efficiency of model reduction methods: ideally, the approximation errors resulting from the reduction should be computed \emph{without} executing the original system. 

Thus, in this paper we propose an approximation method for biochemical networks, in which no prior knowledge about the original system's behavior is required. Our method combines the dominance concept with mass invariants of the original ODE system in order to compute inequality constraints on the species' concentrations. These constraints are then combined with the original system of equations, in order to obtain a reduced system of ODEs that provides time-dependent lower and upper bounds on the species' concentrations. Depending on the coarseness of detail we choose to incorporate in the mass invariant-generated inequalities, our approach can serve either as a reduction method, or to quantify the approximation errors of tropicalization reduction heuristics. 

To achieve this, we abstract the original system by a box, the hyper-faces of which provide lower and upper bounds for the concentrations of the species. The two equations of the hyper-faces of a species represent simplified versions of the original differential equation of the species, in which only the dominant positive and negative monomials are considered. We refer to these equations as being \emph{tropicalized}. Then, instead of interpreting the differential equations over the state of the original system, we will lift this interpretation conservatively over each hyper-face of the box. To do this, we will bound, for every hyper-face, the derivative of the corresponding coordinate in the solution of the original differential equation over the whole hyper-face. Our method should allow for formal evaluation of tropicalization approaches, and as such the bounds are derived using the dominance relations between monomials of the original ODE. Mass invariants of the original system will then be used to refine the bounds, and thus increase the accuracy of our method. By construction, the maximal solutions of the original, respectively tropicalized (\emph{i.e.}, abstracted) equations are related by the following soundness criterion: when both defined at time \emph{t}, the state of the original system is within the hyper-box of the abstract system. 

\begin{example}\label{ex:1}
Let us consider the equations (2) of the Michaelis-Menten mechanism, under the QSS assumption: $k_{2}\gg k_{-1}$, \emph{i.e.} $\epsilon \cdot k_{-2}\geq k_{-1} \geq 0$, for an $\epsilon \in [0,1]$. From (\ref{def:dom}), it follows that one can write (by extension): $ k_{2}\succ_{\epsilon} k_{-1}$. Then, we can deduce the following lower and upper bounds (that we call \emph{tropicalized}) on the concentration of $x_2$:

\begin{equation}\label{eq:Mmineq}
\begin{cases}
k_{-1}[E:S] - k_{1}[E][S] &\leq \hspace{8pt} \frac{d[S]}{dt} \hspace{8pt}\leq k_{-1}[E:S] - k_{1}[E][S]\\

k_{2}[E:S] - k_1[E][S]&\leq \hspace{8pt}\frac{d[E]}{dt} \hspace{8pt}\leq (1+\epsilon)k_{2}[E:S]-k_1[E][S]\\

k_1[E][S]-(1+\epsilon)k_{2}[E:S]&\leq \hspace{8pt} \frac{d[E:S]}{dt} \leq k_1[E][S]- k_{2}[E:S]\\

k_2[E:S] &\leq\hspace{8pt} \frac{d[P]}{dt} \hspace{8pt}\leq k_2[E:S]
\end{cases}
\end{equation}

For convenience purposes, we will use the notation $x_1,x_2,x_3,x_4$ for the species' concentrations, $[S],[E],[E:S],[P]$. We propose to approximate the state of the system by a box of $\mathbb{R}^4$. A box of $\mathbb{R}^4$ is a set of the form $\{(x_1,x_2,x_3,x_4)\mid \underline{x_i}\leq x_i \leq \overline{x_i},\forall 1\leq i\leq 4\}$, where $(\underline{x_i},\overline{x_i})$ are pairs of numbers satisfying $\underline{x_i}\leq \overline{x_i}$, $\forall 1\leq i \leq 4$. Intuitively,  the real number $\underline{x_i}$ provides a lower bound to the value of the variable $x_i$, and denotes the face $\{(x_1,x_2,x_3,x_4)\in\mathbb{R}^4\mid x_i=\underline{x_i}, \underline{x_j}\leq x_j \leq \overline{x_j}, \forall 1\leq j\leq 4, i\neq j\}$. We will denote this face as $\mathcal{F}_{\underline{x_i}}(\underline{x_1},\overline{x_1},\underline{x_2},\overline{x_2},\underline{x_3},\overline{x_3},\underline{x_4},\overline{x_4})$. The other faces are defined in the same way, and the same reasoning applies to $\overline{x_i}$, which provides an upper bound to the same variable. For ease of notation, we shall use $\underline{\overline{x}}$ to denote the vector $(\underline{x_1},\overline{x_1},\underline{x_2},\overline{x_2},\underline{x_3},\overline{x_3},\underline{x_4},\overline{x_4})$ .

Next, let us consider the following functions:

\begin{equation}
\begin{cases}
F^\#_{\underline{x_1}}(\underline{\overline{x}}) = &k_{-1}  \underline{x_3} - k_1 \overline{x_2} \underline{x_1}\\

F^\#_{\overline{x_1}}(\underline{\overline{x}}) = 
&k_{-1} {\overline{x_3}} - k_1 \underline{x_2} \overline{x_1}\\

F^\#_{\underline{x_2}}(\underline{\overline{x}}) =
&k_{2} \underline{x_3} - k_1 \underline{x_2} \overline{x_1}\\

F^\#_{\overline{x_2}}(\underline{\overline{x}}) = 
&(1+\epsilon)  k_{2} \overline{x_3} - k_1 \overline{x_2} \underline{x_1}\\

F^\#_{\underline{x_3}}(\underline{\overline{x}}) =
& k_1 \underline{x_1} \underline{x_2} - (1+\epsilon)  k_{2} \underline{x_3}\\

F^\#_{\overline{x_3}}(\underline{\overline{x}}) =
& k_1 \overline{x_1} \overline{x_2}-k_{2} \overline{x_3}\\

F^\#_{\underline{x_4}}(\underline{\overline{x}}) =
& k_2  \underline{x_3}\\

F^\#_{\overline{x_4}}(\underline{\overline{x}}) =
& k_2 \overline{x_3}\\

\end{cases}
\end{equation}

The abstraction of the concrete system of equations is then defined as 
$\begin{cases}
\frac{d\underline{x_i}}{dt}=F^\#_{\underline{x_i}}(\underline{\overline{x}})\\
\frac{d\overline{x_i}}{dt}=F^\#_{\overline{x_i}}(\underline{\overline{x}})
\end{cases}$, 

$\forall 1\leq i\leq 4$. 

If we fix the same initial conditions for both the concrete and the abstracted system, $\underline{x_i}(0)=\overline{x_i}(0)=x_i(0), \forall 1\leq i\leq 4$, we can relate the solution of the abstract system to that of the original one. For every $1\leq i\leq 4$, the real number $F^\#_{\underline{x_i}}(\underline{x_1},\overline{x_1},\underline{x_2},\overline{x_2},\underline{x_3},\overline{x_3},\underline{x_4},\overline{x_4})$ provides a lower bound to the value of the function $\frac{dx_i}{dt}$ over the face $\mathcal{F}_{\underline{x_i}}$, whereas the real number $F^\#_{\overline{x_i}}(\underline{x_1},\overline{x_1},\underline{x_2},\overline{x_2},\underline{x_3},\overline{x_3},\underline{x_4},\overline{x_4})$ provides an upper bound to the value of the function $\frac{dx_i}{dt}$ over the face $\mathcal{F}_{\overline{x_i}}$. That is to say, we have $\frac{d\underline{x_i}}{dt}\leq \frac{dx_i}{dt}$, for every pair $(x_1,x_2,x_3,x_4) \in \mathcal{F}_{\underline{x_i}}$, and $\frac{dx_i}{dt}\leq\frac{d\overline{x_i}}{dt}$, for every pair $(x_1,x_2,x_3,x_4)\in \mathcal{F}_{\overline{x_i}}$. Then, using the results of \cite{kirk}, we can conclude that, for every time point $t$, and $\forall 1\leq i \leq 4$, the bounds:

\begin{equation}\label{eq:bounds}
\underline{x_i}(t)\leq x_i(t)\leq \overline{x_i}(t)
\end{equation}

are satisfied. Thus, the solution of the abstract system of equations provides lower and upper bounds for the value of the variables of the original system of equations. 
\end{example}

\begin{figure}\centering
 \includegraphics[width=0.6\columnwidth, height=5cm]{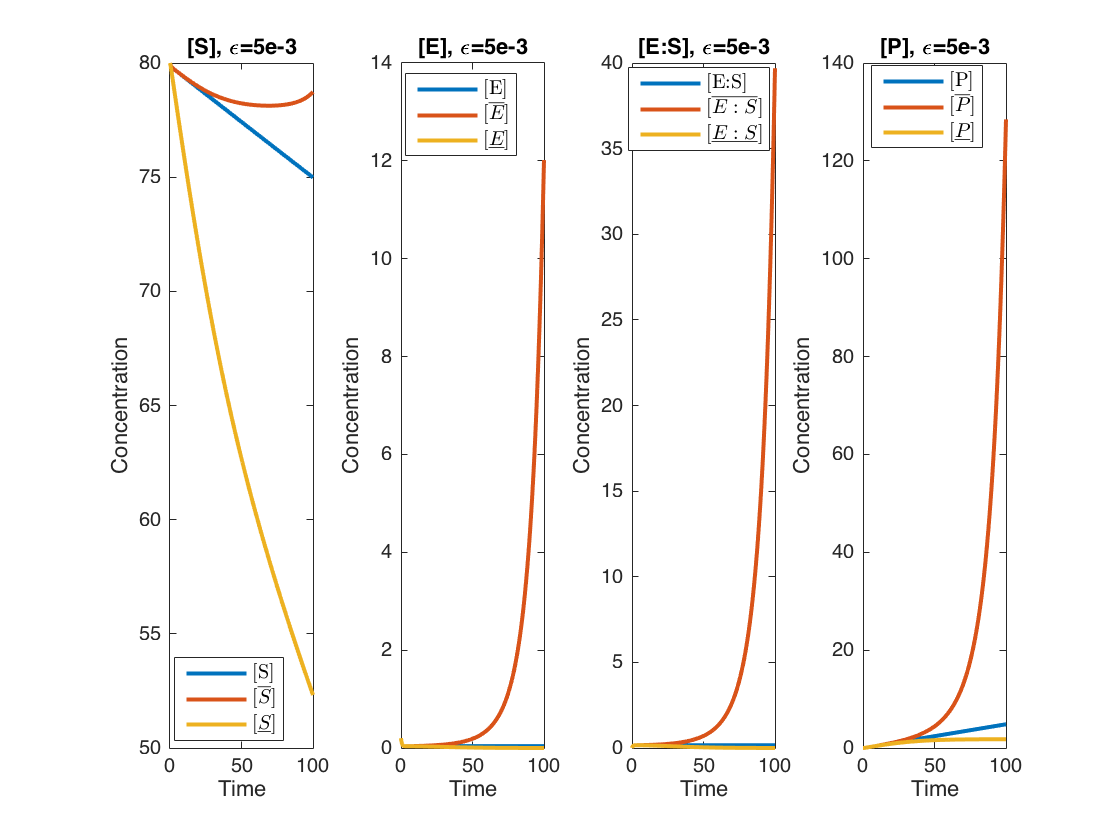}
 \caption{Bounds on the species' concentration with respect to simulation time, with $\epsilon=5\cdot10^{-3}$, rate constants $k_1=0.017, k_{-1}=0.0017, k_2=0.3$, and initial concentrations $[S]=80,[E]=0.2,[E:S]=0,[P]=0$. For each of the 4 species, $[\underline{\cdot }]$ and $[\overline{\cdot }]$ denote the lower, respectively upper bounds on its concentration. The depicted results were obtained without using the mass invariants of the original system to constraint the bounds, and as such are suboptimal. }  
\label{fig:MMunrefined}
\end{figure}

\begin{remark}\label{note1} In the above example, in order to obtain safe lower/upper bounds on $x_i$'s concentration, we make the variables range over the hyper-faces. One notices that the variable $x_i$ is treated specifically in the derivatives of the variables $\underline{x_i}, \overline{x_i}$ - any of its occurences is replaced by the variable corresponding to the hyper-face we want to bound. By contrast, the other variables, $x_j$, are replaced according to the sign of their occurence:

\begin{itemize}

\item in $\frac{d\underline{x_i}}{dt}$, $x_j$ is replaced with $\begin{cases}
\overline{x_j}, \text{ if } x_j \text{ occurs negatively, } \\
\underline{x_j}, \text{ if } x_j \text{ occurs positively. }
\end{cases}$

 \item in $\frac{d\overline{x_i}}{dt}$, $x_j$ is replaced with $\begin{cases}
\overline{x_j}, \text{ if } x_j \text{ occurs positively, } \\
\underline{x_j}, \text{ if } x_j \text{ occurs negatively. }
\end{cases}$
\end{itemize}

This comes from the fact that the derivative on $x_i$ is evaluated on the corresponding hyper-face, which allows for greatly reducing the loss of precision. For a formal proof of the soundness of this approach, the reader is referred to \cite{kirk}. Intuitively, it is justified by the intermediate value theorem: given a family of functions $\{f_i\}$ over the real field, if one function $f_i$ does not take the highest value at time $t$, whereas it is the case at time $t'' > t$, then necessarily, there exists a time $t'$ such that $t < t' \leq t''$ in which $f_i$ takes the highest value while crossing another function of the family. 

\end{remark}

In Example \ref{ex:1}, the inequality constraints on the concentrations of species were determined based on reaction rates constants that verify the QSS condition. In Fig.\ref{fig:MMunrefined}, we show the time-evolution of the bounds on the concentration of the 4 species in the Michaelis-Menten system, for an arbitrarily chosen set of reaction rate constants and initial concentrations that satisfy the QSS condition (\emph{i.e.}, $k_{2}\succ_\epsilon k_{-1}$, and $[S]\succ_\epsilon [E]+[E:S]$ at time $t=0$). Nonetheless, our model reduction is sound no matter the value of initial concentrations and reaction rate constants. The equations have been integrated using the solver $ode15s$ of Matlab\cite{matlab}. Strictly speaking, numerical errors stemming from numerical integration may accumulate throughout the simulation, but herein we choose to ignore them.

In Fig.\ref{fig:MMunrefined}, we notice that the bounds diverge at a fast rate from the original trajectory, despite the restriction of the derivative's evaluation on the hyper-face of the box (as explained in Note 1). A way to improve accuracy is to take into account the original system's mass invariants, when computing the bounds. In general, a biochemical system can have several conservation laws/mass invariants, which are linear functions $b_1(\textbf{x}),\ldots,b_m(\textbf{x})$ of the concentrations, that are constant in time. These equality constraints can be used to refine the bounds on the initial system's species' concentrations. 
We can safely further restrict the evaluation of the derivative of each coordinate to the intersection of the corresponding hyperface with the subspace delimited by the conservation laws containing the variable itself. Because a variable can appear in more than one mass invariant, we choose to keep the most optimistic bound that can be computed by intersecting the hyper-face with the mass invariant subspace: the greatest lower bound, respectively the smallest upper bound. 

\begin{example}\label{ex:2}
In the  Michaelis-Menten system, the total number of enzymes is constant, and so is the overall number of substrates and product. The two conservation laws can be written as:
$\begin{cases}
x_2(t) + x_3(t) = e_0\\
x_1(t) + x_3(t) + x_4(t) = s_0
\end{cases}$, 
with $e_0=x_2(0)+x_3(0)$, and $s_0=x_1(0)+x_3(0)+x_4(0)$.

Assuming once more that $k_{2}\gg k_{-1}$, by substituting $x_3$ by $e_0-x_2$ or $s_0-x_1-x_4$ into \ref{eq:Mmineq}, three equivalent tropicalized upper bounds on the concentration of $x_2$ are obtained:

\begin{equation}\label{eq:MMpart}
\begin{cases}
\frac{dx_2}{dt}\leq (1+\epsilon)  k_{2}  x_3 - k_1  x_2  x_1\\

\frac{dx_2}{dt}\leq (1+\epsilon)  k_{2}  (e_0-x_2) - k_1  x_2  x_1\\

\frac{dx_2}{dt}\leq (1+\epsilon)  k_{2}  (s_0-x_1-x_4) - k_1  x_2  x_1 \leq (1+\epsilon)  k_{2}  (s_0-x_1) - k_1  x_2  x_1 
\end{cases}
\end{equation}

Lifting the interpretation of the differential equations over the hyper-face corresponding to $\overline{x_2}$ results in three different expressions for the upper bound on $\frac{dx_2}{dt}$, of possibly different accuracies:

\begin{equation}
\begin{cases}
\frac{d{\overline{x_{2,1}}}}{dt} = (1+\epsilon)  k_{2}  \overline{x_3} - k_1  \overline{x_2}  \underline{x_1}\\

\frac{d{\overline{x_{2,2}}}}{dt} = (1+\epsilon)  k_{2}  (e_0-\overline{x_2}) - k_1  \overline{x_2}  \underline{x_1}\\

\frac{d{\overline{x_{2,3}}}}{dt} = (1+\epsilon)  k_{2}  (s_0-\underline{x_1}) - k_1  \overline{x_2}  \underline{x_1}\\
\end{cases}
\end{equation}

The most accurate sound upper bound on $\frac{dx_2}{dt}$ then writes as:
\begin{equation}
 \min(\frac{d{\overline{x_{2,1}}}}{dt},\frac{d{\overline{x_{2,2}}}}{dt},\frac{d{\overline{x_{2,3}}}}{dt}) = (1+\epsilon) \cdot k_{2} \cdot \min(\overline{x_3},e_0-\overline{x_2},s_0-\underline{x_1}) - k_1  \overline{x_2}  \underline{x_1}
\end{equation}

\begin{note}
The choice to introduce $\min$ and $\max$ operations in the expressions of the computed bounds is accounted for by our initial motivation: because existing tropicalization reduction heuristics are  not justified by rigourous estimates, we aim to provide a method for \emph{quantifying errors} stemming from such tropicalization reduction approaches, at the same time creating a tropicalization approach with \emph{guarantees}\footnote{We nonetheless stress that our goal is not to correct the faults of existing tropicalization-inspired reduction methods, but rather quantify them by proposing a more rigorous tropicalization approach, in which the dominated monomials are bounded, rather than discarded from the ODEs}. As such, we aim at computing error bounds that are as precise as possible, hence the choice of using $\min$ and $\max$ operations for bound refinement, albeit with the disadvantage of using functions that are not $\mathbb{C}^1$, thus introducing non-smooth vector fields. The trade-off between smoothness and precision can be tuned according to the desired goal: less precise bounds can be obtained by choosing to use smooth functions. Moreover, smoothness of vector fields is generally not guaranteed during the numerical simulation of biochemical models: as the model variables represent biochemical species' concentrations, a good practice is to call the numerical solvers used to approximate the system's behavior using with the 'Non-Negative' option, which amounts to introducing a $\max$ operation into the equations (\emph{i.e.}, $\max(0,x_i)$),  in order to prevent negative values of variables.
\end{note}
The same reasoning can be applied to all variables appearing in the expression of $\frac{dx_2}{dt}$, in order to obtain the most accurate upper bound: 

\begin{equation}\label{eq:pre}
\frac{d\overline{x_2}}{dt} = (1+\epsilon) \cdot k_{2} \cdot \min(\overline{x_3},e_0-\overline{x_2},s_0-\underline{x_1}) -
k_1\cdot\max(\underline{x_1},0)\cdot\max (\overline{x_2},e_0-\overline{x_3})
\end{equation}
 
\end{example}

\begin{figure}
\centering
\begin{subfigure}{.5\textwidth}
  \centering
  \includegraphics[width=\linewidth]{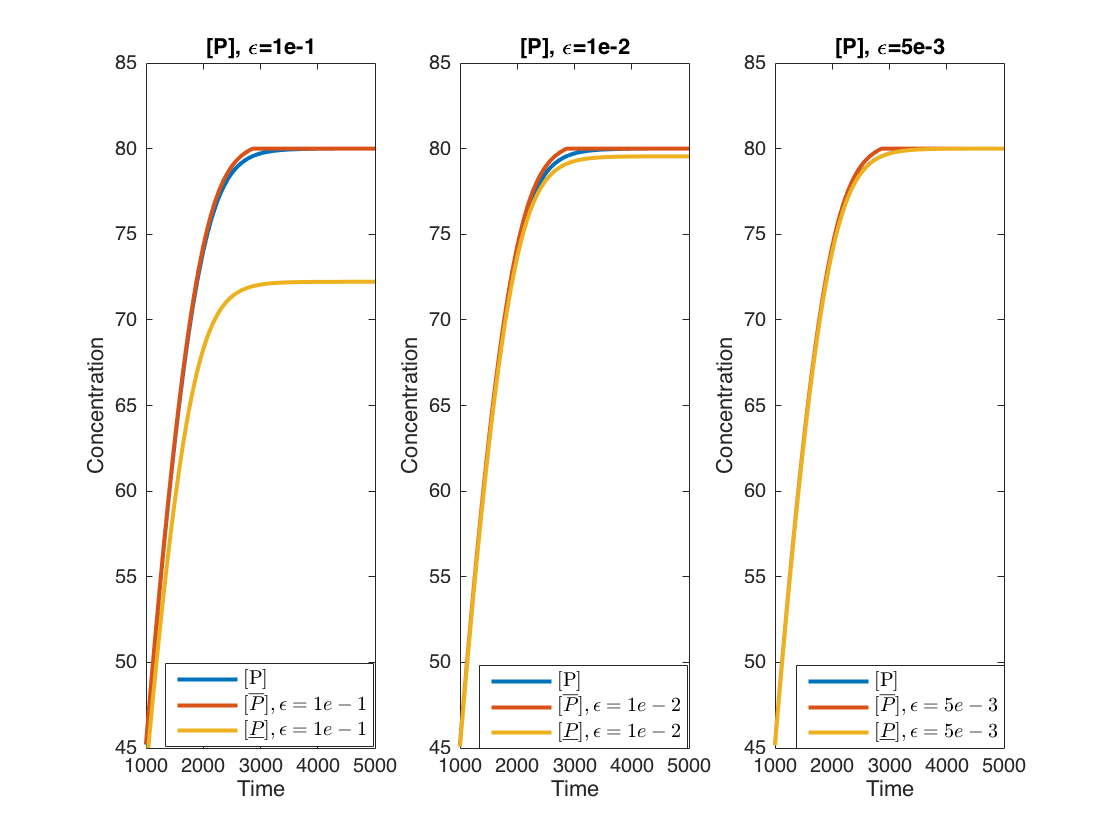}
  \label{fig:MMrefined}
\end{subfigure}%
\begin{subfigure}{.5\textwidth}
  \centering
  \includegraphics[width=\linewidth]{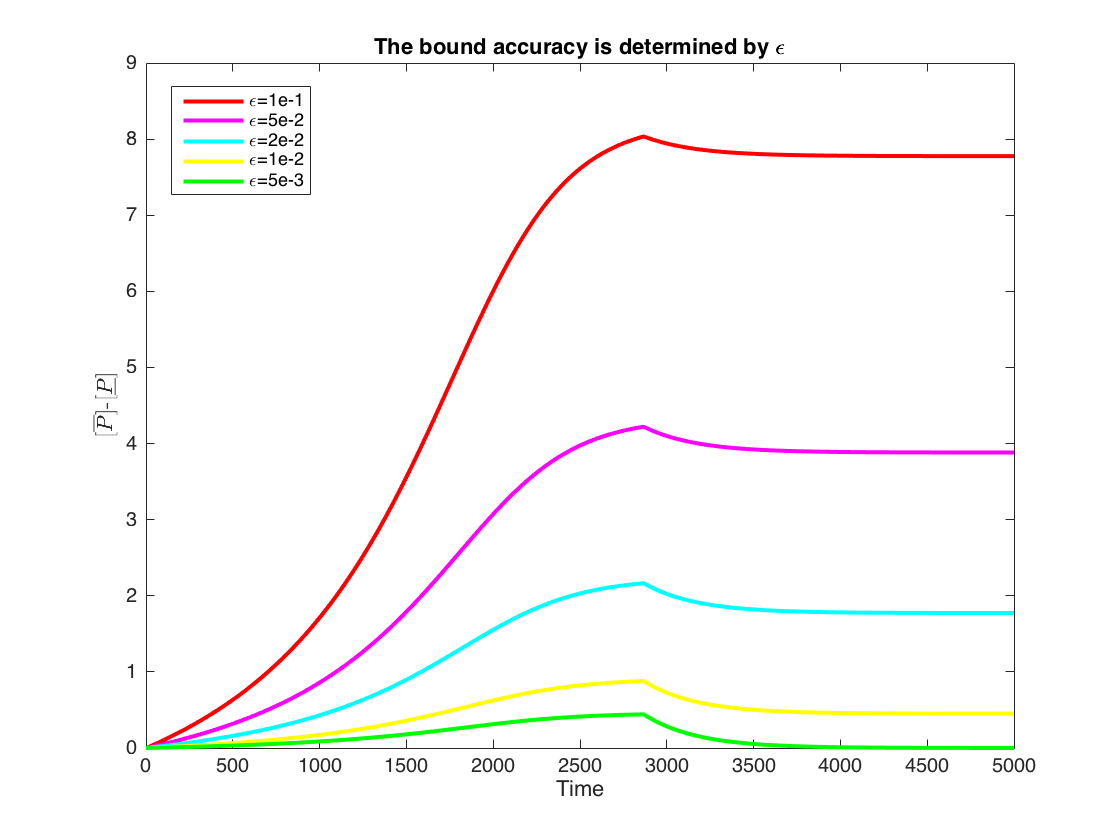}
  \label{fig:bacc}
\end{subfigure}
\caption{\emph{Left}: bounds on the concentration of P, obtained by simulating the ODE system in Example\ref{eq:redrefine}, for different values of $\epsilon$.  Rate constants and initial concentration as in Fig.\ref{fig:MMunrefined}: $k_1=0.017, k_{-1}=0.0017, k_2=0.3,[S]=80,[E]=0.2,[E:S]=0,[P]=0$. \emph{Right}: For different values of $\epsilon$, the accuracy of the resulting bounds is computed as the difference between the upper and the lower bound.}
\label{fig:MMrefinedbacc}
\end{figure}




\begin{note}
In (\ref{eq:MMpart}), when computing the third bound, instead of substituting $x_3$ by its conservation law expression, $s_0-x_1-x_4$, we choose to bound its value by an expression not containing $x_4$. We do so in order to avoid introducing supplementary variables w.r.t. those present in the tropicalized original bound (\emph{i.e.}, $x_1$, $x_2$ and $x_3$). This method, in which mass invariant partial refinement is introduced after the tropicalized bounds have been computed, can be considered as a \emph{per se} model reduction method, as no supplementary information/complexity is introduced by incorporating the conservation laws. By contrast, the approach in which all the information contained by the conservation laws is exploited in order to derive the most accurate bounds can constitute a method of error-estimation for tropicalization based reduction heuristics. We present the two different methods formally in Section 3.
\end{note}

The issue of specifying QSS species and QE reactions a priori, when performing model reductions, is circumvented by our method. Instead, the notion of \emph{region} is used in order to eliminate monomials from the species' ODEs. Our method uses static inspection of each ODE, in order to partition the state space into different regions according to which production, respectively consumption terms dominate the others. Using this partitioning, simplified expressions bounding the species concentrations are derived for each region of the state space, allowing symbolic simplification and limiting numerical approximations.

\begin{example} In the case of the Michaelis-Menten mechanism, there are three possible dominance regions for $\frac{dx_2}{dt}$ leading to three possible pairs of lower and upper bounds:
\begin{enumerate}
\item Region 1, if $k_{-1}$ dominates $k_2$: 

 $\epsilon k_{-1} \geq k_2 \Rightarrow  k_{-1}  x_3-k_1 x_2 x_1\leq\frac{dx_2}{dt}\leq (1+\epsilon) k_{-1}  x_3-k_1  x_2   x_1$

\item Region 2, if $k_{2}$ dominates $k_{-1}$: 

$\epsilon k_2 \geq k_{-1} \Rightarrow  k_2  x_3-k_1  x_2   x_1 \leq\frac{dx_2}{dt}\leq (1+\epsilon)  k_2  x_3-k_1  x_2  x_1$

\item Region 3, if there is no dominant rate (\emph{i.e.}, $k_{-1}$ and $k_2$ are of comparable magnitude):

$\begin{cases}\epsilon k_{-1}\leq k_2\\ \epsilon k_{2}\leq k_{-1}\end{cases} \Rightarrow (k_{-1}+k_2)  x_3 -k_1  x_2   x_1 \leq \frac{dx_2}{dt} \leq (k_{-1}+k_2)  x_3-k_1  x_2  x_1$
\end{enumerate}

\end{example}

The complete system of equations obtained using mass invariants refinement of bounds, for all the possible dominance regions, can be found in Example \ref{eq:redrefine}. The improvement of bound accuracy via mass invariants can be observed in Fig. \ref{fig:MMrefinedbacc}. As expected, one can also observe in Fig.\ref{fig:MMrefinedbacc} that results become more precise as the value of $\epsilon$ increases, \emph{i.e.} as $k_{-1}$ and $k_2$ become more separated. 

\begin{example}\label{eq:redrefine} For convenience purposes, denote the species concentrations, $[S],[E],[E:S],[P]$, using $x_1,x_2,x_3,x_4$. Then, the derivatives of the lower and upper bounds of the original system's species' concentrations write as:
\begin{itemize}

\item $\frac{d\underline{x_1}}{dt} = k_{-1}  \max (\underline{x_3},e_0-\overline{x_2}) - k_1  \min(\underline{x_1},s_0-\underline{x_3})  \min(\overline{x_2},e_0-\underline{x_3})$

\item $\frac{d\overline{x_1}}{dt} = k_{-1}  \min(\overline{x_3},e_0-\underline{x_2},s_0-\overline{x_1}) - k_1  \min(\overline{x_1},s_0-\underline{x_3})  \max (\underline{x_2},e_0-\overline{x_3})$

\item $\frac{d\underline{x_2}}{dt} = \underline{c_+}   \max (\underline{x_3},e_0-\underline{x_2})-
k_1  \min(\overline{x_1},s_0-\underline{x_3})  \min(\underline{x_2},e_0-\underline{x_3})$, with 
$\underline{c_+}=\begin{cases}
k_{-1}, &\text{ if } \epsilon  k_{-1}\geq k_2\\
k_2, &\text{ if } \epsilon  k_2\geq k_{-1} \\
(k_{-1}+k_2), &\text{ otherwise }
\end{cases}$


 

\item $\frac{d\overline{x_2}}{dt} = \overline{c_+}  \min(\overline{x_3},e_0-\overline{x_2},s_0-\underline{x_1}) -
k_1  \max (\underline{x_1},0)   \max (\overline{x_2},e_0-\overline{x_3})$, with $\overline{c_+}=
\begin{cases}
(1+\epsilon)  k_{-1}, &\text{ if } \epsilon  k_{-1}\geq k_2 \\
(1+\epsilon)  k_2, &\text{ if } \epsilon  k_2\geq k_{-1} \\
(k_{-1}+k_2), &\text{ otherwise }
\end{cases}$

              
              

\item $\frac{d\underline{x_3}}{dt} = k_1  \max (\underline{x_1},0)  \max (\underline{x_2},e_0-\underline{x_3}) - \underline{c_-}  \min(\underline{x_3},e_0-\underline{x_2},s_0-\underline{x_1})$, with $\underline{c_-}=
\begin{cases}
(1+\epsilon)  k_{-1}, &\text{ if } \epsilon  k_{-1}\geq k_2 \\
(1+\epsilon)  k_2, &\text{ if } \epsilon  k_2\geq k_{-1} \\
(k_{-1}+k_2), &\text{ otherwise }
\end{cases}$

\item $\frac{d\overline{x_3}}{dt} =  k_1  \min(\overline{x_1},s_0-\overline{x_3})  \min(\overline{x_2},e_0-\overline{x_3}) - \overline{c_-}  \max (\overline{x_3},e_0-\overline{x_2})$,with $\overline{c_-}=
\begin{cases}
 k_{-1}, &\text{ if } \epsilon  k_{-1}\geq k_2 \\
 k_2, &\text{ if } \epsilon  k_2\geq k_{-1} \\
(k_{-1}+k_2), &\text{ otherwise }
\end{cases}$

\item $\frac{d\underline{x_4}}{dt} = k_2  \max (\underline{x_3},0)$
\item $\frac{d\overline{x_4}}{dt} = k_2  \min(\overline{x_3},s_0-\overline{x_4})$
\end{itemize}

\end{example}

The Michaelis-Menten system represents a particular, simple case study: the choice of reaction rate constants fixes the dominance region in which the system evolves. In general, the state of a biochemical network can traverse multiple such regions, as the dominant monomials can change from one concentration domain to another. Thus, we next introduce a case study in which the dominant monomials are concentration-dependent, which in turn means that the dominance region is no longer fixed. Our method is designed with this more general situation in mind: having computed the most accurate bounds for each region of the state space partitioning, and having no information regarding the region in which the original system evolves at a given time $t$, our approach chooses the least accurate local bound, in order to ensure global soundness. 

\subsection{Motivating example: A DNA model}\label{sub:DNA}

We construct a simple extension of the Michaelis-Menten system, in which the product formation reaction is catalazyed by a dimer of an enzyme $M$. The reaction system and its ODE system\footnote{once again, we denote the species $M,M_2,DNA, M_2.DNA,P$ by $x_1,x_2,x_3,x_4,x_5$} read:

\begin{minipage}{.45\linewidth}
\begin{equation*}
\begin{cases}
M + M \underset{k_{-1}}{\stackrel{k_1}{\rightleftharpoons}}  M_2 \\
M_2 + DNA\underset{k_{-2}}{\stackrel{k_2}{\rightleftharpoons}} M_2.DNA \\
M_2.DNA \xrightarrow{k_3} DNA + P
\end{cases}
\end{equation*}
\end{minipage}%
\begin{minipage}{.45\linewidth}


\begin{equation}\label{eq:DNAexample}
\begin{cases}\frac{dx_1}{dt} = -2  k_1  x_1^2 + 2  k_{-1}  x_2\\
\frac{dx_2}{dt}= -k_{-1}  x_2-k_2  x_2  x_3 + k_{-2}  x_4 +k_1  x_1^2\\
\frac{dx_3}{dt}= k_{-2}  x_4 +k_3 x_4-k_2  x_2  x_3 \\
\frac{dx_4}{dt}= -k_{-2}  x_4 - k_3  x_4 + k_2  x_2  x_3\\
\frac{dx_5}{dt} = k_3  x_4
\end{cases}
\end{equation}
\end{minipage}

The mass invariants write:

\begin{equation}\label{eq:DNAmass}
\begin{cases}
 x_1 + 2  x_2 + 2  x_4 + 2  x_5 = M_0 \\ 
 x_3 + x_4 = DNA_0
\end{cases}
\end{equation}

Dominance regions become concentration dependent: for example, the dominant positive monomial in $\frac{dx_2}{dt}$ is determined by the dominance relations between both the concentrations of $x_1$ and $x_4$, and between reaction rate constants $k_1$ and $k_{-2}$.

This DNA example will serve as a case study for the remainder of our paper. 


\section{Combining ODEs and mass invariants}

\subsection{Model reduction using conservative numerical approximations}

The guarantees of our method are a consequence of a carefully designed symbolic propagation of inequality constraints on the species' concentrations. Thus, symbolic transformations have to be applied on numerical expressions, of which we introduce a syntax and semantics. We also introduce an alternative definition of a biochemical model to that presented in Sect. 2, which is then used to define and justify our approximation method. 

\begin{definition}\label{def:syn}\textbf{(Syntax of expressions)} Let $\mathcal{S}$ be a set of variables. We define an $\mathcal{S}$-expression inductively, as follows\footnote{the syntactic operators are written using a superscript dot, in order to distinguish them from their associated mathematical functions}:
\begin{enumerate}
\item each positive real number $k\in \mathbb{R}_+$ is an $\mathcal{S}$-expression; 
\item each variable $x \in \mathcal{S}$ is an $\mathcal{S}$-expression; 
\item if $e$ is an $\mathcal{S}$-expression, then $(\dot{-}e)$ is an $\mathcal{S}$-expression; 
\item if $e_1$ and $e_2$ are $\mathcal{S}$-expressions, then $(e_1\dot{+}e_2)$, $(e_1\dot{\cdot }e_2)$, $\dot{\min}(e_1,e_2)$, $\dot\max (e_1,e_2)$ are all $\mathcal{S}$-expressions;
\end{enumerate}
The set of $\mathcal{S}$-expressions is denoted as $Expr_\mathcal{S}$. Given an $\mathcal{S}$-expression $e$, we define its \emph{support}, denoted $supp(e)$, as the set of variables it contains.
\end{definition}



\begin{definition}\label{def:sem}\textbf{(Semantics of expressions)} Let $\mathcal{S}$ be a set of variables and $e$ be an $\mathcal{S}$-expression. The semantics of the expression $e$ is the function $\llbracket e\rrbracket_\mathcal{S}:\mathbb{R}^\mathcal{S}\rightarrow\mathbb{R}$, defined inductively as follows:
\begin{enumerate}

\item $\forall c \in \mathbb{R}, \llbracket c\rrbracket_{\mathcal{S}(\rho)} = c$

\item $\forall x\in\mathcal{S}, \llbracket x \rrbracket_{\mathcal{S}(\rho)} = \rho(x)$

\item $\forall e\in Expr_{\mathcal{S}}, \llbracket\dot{-}e\rrbracket_{\mathcal{S}(\rho)}=-\llbracket e \rrbracket_{\mathcal{S}(\rho)}$

\item $\forall e_1, e_2 \in Expr_{\mathcal{S}}, \llbracket e_1\dot{+}e_2\rrbracket_{\mathcal{S}(\rho)}=\llbracket e_1 \rrbracket_{\mathcal{S}(\rho)}+\llbracket e_2 \rrbracket_{\mathcal{S}(\rho)}$

\item $\forall e_1, e_2 \in Expr_{\mathcal{S}}, \llbracket e_1\dot{\cdot }e_2\rrbracket_{\mathcal{S}(\rho)}=\llbracket e_1 \rrbracket_{\mathcal{S}(\rho)} \llbracket e_2 \rrbracket_{\mathcal{S}(\rho)}$

\item $\forall e_1, e_2 \in Expr_{\mathcal{S}}, \llbracket \dot\min(e_1,e_2)\rrbracket_{\mathcal{S}(\rho)}=\min(\llbracket e_1 \rrbracket_{\mathcal{S}(\rho)}\llbracket e_2 \rrbracket_{\mathcal{S}(\rho)})$

\item $\forall e_1, e_2 \in Expr_{\mathcal{S}}, \llbracket \dot\max (e_1,e_2)\rrbracket_{\mathcal{S}(\rho)}=max (\llbracket e_1 \rrbracket_{\mathcal{S}(\rho)}\llbracket e_2 \rrbracket_{\mathcal{S}(\rho)})$
\end{enumerate}
for every environment $\rho \in \mathbb{R}^{\mathcal{S}}$.
\end{definition}

We use Defs. \ref{def:syn} and \ref{def:sem} to define the notion of system of symbolic differential equations and symbolic equality constraints derived from conservation laws. 

\begin{definition}\textbf{(Symbolic ODE system)} 

A system of symbolic ordinary differential equations and equality constraints modeling a biochemical network is a tuple $(\mathcal{S},\mathbb{I},\mathbb{F},(\mathbb{E}_b))$, where:
	\begin{itemize}
    	\item $\mathcal{S}=\{x_1,\ldots,x_s\}$ is a set of variables, denoting species' concentrations, 
        \item $\mathbb{I}:\mathcal{S}\rightarrow\mathbb{R}^+$ is a non-negative function, mapping each species to its initial concentration,
        \item $\mathbb{F}:\mathcal{S}\rightarrow Expr_\mathcal{S}$ is a function describing the evolution of species' concentrations, as described in Eq.(\ref{eq:ODE}):
        \begin{equation*}
          \forall x_i \in \mathcal{S}, \mathbb{F}(x_i) = P_i^+(\textbf{x}) - P_i^-(\textbf{x}),
        \end{equation*}
        with  $P_i^{+/-} \in Expr_{\mathcal{S}}$, Laurent polynomials with positive coefficients,
        
        \item $\{\mathbb{E}_b\}$\footnote{the number $b$ indexes the different ways of expressing a species $x_i$, by using the mass invariants in which it appears} is a family of functions from the set $\mathcal{S}$ into the set $Expr_\mathcal{S}$, denoting equality constraints derived from conservation laws, such that $\forall f:\mathcal{S}\rightarrow\mathbb{R}^{\mathbb{R}^+}$ satisfying 
        \begin{equation*}
        \begin{cases}
        f(x_i)(0) = \mathbb{I}(0), &\forall x_i\in\mathcal{S}\\
        \frac{df(x_i)}{dt}(t)=\llbracket\mathbb{F}(x_i)\rrbracket_{\mathcal{S}[x_i\mapsto f(x_i)(t)]}, &\forall x_i\in\mathcal{S} \text{ and } t\in\mathbb{R}^+ 
        \end{cases}
        \end{equation*}
    the constraint 
        \begin{equation*}
         f(x_i)(t) = \llbracket\mathbb{E}_b(x_i)\rrbracket_{\mathcal{S}[x_i\mapsto f(x_i)(t)]}
        \end{equation*}
        
       is satisfied for every function $\mathbb{E}_b$ of the family $\{\mathbb{E}_b\}$, $\forall x_i\in\mathcal{S}$, and for every time $t\in\mathbb{R}^+$.
        \end{itemize} 
\end{definition}

\begin{example}\label{ex:def}\textbf{(A DNA example)} In our running example, $\mathcal{S}=\{x_1,x_2,x_3,x_4\}$, $\mathbb{F}$ is defined by the equations of (\ref{eq:DNAexample}), and the equality constraints derived from the conservation laws of (\ref{eq:DNAmass}) write:
\begin{equation}\label{eq:eq-const}
\begin{cases}
\mathbb{E}_1(x_1)= M_0-2 x_2 - 2 x_4 - 2 x_5; &\mathbb{E}_2(x_1)= M_0 - 2 DNA_0 - 2 x_2 + 2 x_4 - 2 x_5 \\
\mathbb{E}_1(x_2)= \frac{M_0-x_1}{2}- x_4- x_5; & \mathbb{E}_2(x_2)= \frac{M_0-x_1}{2}-DNA_0+x_3- x_5 \\

\mathbb{E}_1(x_3)= DNA_0 - x_4; &\mathbb{E}_2(x_3) = DNA_0 -\frac{M_0-x_1}{2}+x_2+x_5;\\

\mathbb{E}_1(x_4) = DNA_0-x_3; &\mathbb{E}_2(x_4)=\frac{M_0-x_1}{2} - x_2 - x_5 \\

\mathbb{E}_1(x_5)= \frac{M_0-x_1}{2}- x_2- x_4; & \mathbb{E}_2(x_5)= \frac{M_0-x_1}{2}-DNA_0+x_3- x_2 \\
\end{cases}
\end{equation}
\end{example}

We partition the state space of each ODE into regions, each one defined by the corresponding pair of dominant monomials, ($Dom(P_i^+(\textbf{x})),Dom(P_i^-(\textbf{x}))$). At any given time $t$, several monomials can be dominant, which can lead to an exponential number of possible regions. To circumvent this issue and obtain a linear number of regions, we choose to replace each region that has more than one dominant term with the unique region in which no term is dominant: if $| Dom(P_i^\pm(\textbf{x}))| > 1$, we choose to keep $P_i^\pm(\textbf{x})$ in the reduced ODE, instead of replacing it with $Dom(P_i^\pm(\textbf{x}))$. The following definition formalizes these concepts.

\begin{definition}\label{def:trop}\textbf{(State partitioning of a symbolic ODE)} Let $(\mathcal{S},\mathbb{I},\mathbb{F},\{\mathbb{E}_b)\}$ be a symbolic ODE system, and $\epsilon\in [0,1]$ a scale separation constant. Then, for every variable $x_i\in\mathcal{S}$, if $P_i^+=\sum\limits_{j=1}^p M_j^+$ and $P_i^-=\sum\limits_{j=1}^n M_j^-$, its state space can be partitioned into $(p+1)\times (n+1)$ regions, each one determined by the corresponding pair of dominant monomials 

\begin{equation}\label{eq:part}
r_i^{k,l} := \begin{cases}
(M_k^+,M_l^-), &\text{ if }  k\leq p, l\leq n, Dom(P_i^+)=M_k^+, Dom(P_i^-)=M_l^-,   \\
(P_i^+, M_l^-), &\text{ if } k=p+1, l\leq n, Dom(P_i^-)=M_l^-   \\
(M_k^+,P_i^-), &\text{ if } k\leq p, l=n+1, Dom(P_i^+)=M_k^+ \\
(P_i^+,P_i^-), &\text{ if } k=p+1, l=n+1
\end{cases}
\end{equation}
\end{definition}

\begin{example}\textbf{(A DNA example)} In Eq.(\ref{eq:DNAexample}), the state space of $x_2$ can be partitioned in 9 regions, as its ODE contains 2 positive terms and 2 negative terms:

$\begin{array}{ccc}
r_2^{1,1} = (k_1 x_1^2 , k_{-1} x_2); \quad r_2^{2,1} = (k_{-2} x_4, k_{-1} x_2); \quad r_2^{3,1} = (k_1 x_1^2+k_{-2} x_4,k_{-1} x_2)\\
r_2^{1,2} = (k_1 x_1^2, k_2  x_2  x_3); \quad r_2^{2,2} = (k_{-2}  x_4,  k_2  x_2  x_3); \quad r_2^{3,2} = (k_1  x_1^2+k_{-2}  x_4,k_2  x_2  x_3)\\
r_2^{1,3} = (k_1  x_1^2, k_{-1}  x_2+k_2  x_2  x_3); \quad r_2^{2,3} = (k_{-2}  x_4, k_{-1}  x_2+k_2  x_2  x_3); \quad r_2^{3,3} = (k_1  x_1^2+k_{-2}  x_4 , k_{-1} x_2+k_2  x_2  x_3)
\end{array}$

\end{example}

We next use the dominance relations that define each region, in order to obtain region-specific lower and upper bounds on the ODE being considered. The next definition formalizes this procedure: 

\begin{definition}\label{def:bounds}\textbf{(Region-specific tropicalized bounds)} Given a symbolic ODE system $(\mathcal{S},\mathbb{I},\mathbb{F},(\mathbb{E}_b))$, and the set of regions $r_i^{k,l}$ for each species $x_i$, the dominance definition \ref{def:dom} can be used to define the following functions, for every region $r_i^{k,l}$:
\begin{equation*}
\mathbb{F}^{k,l}_\downarrow(x_i):=\begin{cases}
M_k^+ \dot{-} (1\dot{+}(n\dot{-}1)\dot\cdot \epsilon)\dot\cdot  M_l^-, &\text{ if } k\leq p, l\leq n, Dom(P_i^+)=M_k^+, Dom(P_i^-)=M_l^-  \\
P_i^+ \dot- (1\dot+(n\dot-1)\dot\cdot \epsilon)\dot\cdot  M_l^-, &\text{ if } k=p+1, l\leq n, Dom(P_i^-)=M_l^- \\
M_k^+ \dot- P_i^-, &\text{ if } k\leq p, l=n+1, Dom(P_i^+)=M_k^+ \\
P_i^+ \dot- P_i-, &\text{ if } k=p+1, l=n+1
\end{cases}
\end{equation*}

\begin{equation*}
\mathbb{F}^{k,l}_\uparrow(x_i) := \begin{cases}
(1\dot+(p\dot-1)\dot\cdot \epsilon)\dot\cdot  M_k^+ \dot-  M_l^-, &\text{ if } k\leq p, l\leq n, Dom(P_i^+)=M_k^+, Dom(P_i^-)=M_l^-  \\
P_i^+ \dot- M_l^-, &\text{ if } k=p+1, l\leq n, Dom(P_i^-)=M_l^- \\
(1\dot+(p\dot-1)\dot\cdot \epsilon)\dot\cdot  M_k^+ \dot- P_i^-, &\text{ if } k\leq p, l=n+1, Dom(P_i^+)=M_k^+ \\
P_i^+ \dot- P_i-, &\text{ if } k=p+1, l=n+1
\end{cases}
\end{equation*}

Functions $\mathbb{F}^{k,l}_\downarrow$ and $\mathbb{F}^{k,l}_\uparrow$ provide symbolic tropicalized lower, resp. upper bounds for $\mathbb{F}(x_i)$ on region $r_i^{k,l}$.  
\end{definition}

\begin{example}\textbf{(A DNA example)}
In our running example, in region $r_2^{2,1}=(k_{-2}x_4,k_{-1}x_2)$, the dominant positive (production) monomial is $k_{-2}x_4$, and the dominant negative (consumption) monomial is $k_{-1}x_2$. Formally, this writes as $\epsilon \cdot  k_{-2}  x_4 \geq k_1  x_1^2 \geq 0$, and $\epsilon \cdot  k_{-1}  x_2 \geq k_2  x_2   x_3 \geq 0$.

Thus, the $r_2^{2,1}$ specific tropicalized bounds write as:
\begin{equation*}\label{eq:trop-bounds}
\begin{cases}
\mathbb{F}_\downarrow^{2,1}(x_2)=k_{-2}  x_4 - (1+\epsilon)  k_{-1}  x_2 \\
 \mathbb{F}_\uparrow^{2,1}(x_2) = (1+\epsilon)  k_{-2}  x_4 - k_{-1}  x_2
\end{cases},
\end{equation*}

which by construction satisfy $\mathbb{F}_\downarrow^{2,1}(x_2)\leq \frac{dx_2}{dt} \leq \mathbb{F}_\uparrow^{2,1}(x_2)$.

\end{example}



The bounds of Def. \ref{def:bounds} can further be refined by using the mass invariants given by the family of functions $\{\mathbb{E}_b\}$, as follows: 

\begin{definition}\label{def:refbounds}\textbf{(Region-specific refined tropicalized bounds)} Given a symbolic ODE system $(\mathcal{S},\mathbb{I},\mathbb{F},(\mathbb{E}_b))$, the set of regions $r_i^{k,l}$ and the symbolic tropicalized bounds $\mathbb{F}^{k,l}_\downarrow(x_i)$, $\mathbb{F}^{k,l}_\uparrow(x_i)$ for each species $x_i$, we define the following bounds: 

\begin{equation*}
\forall r_i^{k,l}, \forall x_j\in \mathcal{V}, \begin{cases}
\mathbb{L}^{k,l}_{i,b}(x_j) := \begin{cases}
\mathbb{E}_b(x_j), &\text{ if } \mathcal{V}=\mathcal{V}_b\\
0, &\text{ otherwise}
\end{cases}\\ 

\mathbb{U}^{k,l}_{i,b}(x_j) :=  \begin{cases}
\mathbb{E}_b(x_j), &\text{ if } \mathcal{V}=\mathcal{V}_b\\
\llbracket{E}_b(x_j)\rrbracket_{\mathcal{V}_b\setminus\mathcal{V}[x_j \mapsto b_i^{k,l}(x_j)]} , &\text{ otherwise }
\end{cases}
\end{cases}
\end{equation*} 
with $\mathcal{V}=supp(\mathbb{F}^{k,l}_\downarrow(x_i))=supp(\mathbb{F}^{k,l}_\uparrow(x_i))$, $\mathcal{V}_b=supp({E}_b(x_j))$, for each function $\mathbb{E}_b$ of the family ($\mathbb{E}_b$) that applies to the variable $x_j$, and $b_i^{k,l}(x_j)\in Expr_{\mathcal{V}}$ is either 0, or a bound generated by the dominating monomial inequality constraints.
\end{definition}

 \begin{example}\label{ex:DNAref}\textbf{(A DNA example)} 


 
When dealing with the tropicalized bounds of Ex.(\ref{eq:trop-bounds}), one needs to refine the bounds of the variables in their support: $\mathcal{V}=\{x_2,x_4\}$. We do so by using their respective equality constraints from (\ref{eq:eq-const}): $\mathbb{E}_1(x_2),\mathbb{E}_2(x_2),\mathbb{E}_1(x_4)$, and $\mathbb{E}_2(x_4)$.
 
What's more, the second dominance inequality of region $r_2^{2,1}$ in Ex.(\ref{eq:trop-bounds})can be rewritten as $x_3\leq\epsilon \frac{k_{-1}}{k_2}$. This allows for a new upper bound on variable $x_3$: $b_2^{2,1}(x_3) = \epsilon \frac{k_{-1}}{k_2}\in Expr_{\mathcal{V}}$.

Using Def.\ref{def:refbounds}, the $r_2^{2,1}$-specific bounds on $x_2$ and $x_4$ write as:

$\begin{array}{cccc}
\mathbb{L}_{2,1}^{2,1}(x_2)=0; \quad \mathbb{L}_{2,2}^{2,1}(x_2)=0; \quad \mathbb{L}_{2,1}^{2,1}(x_4) = DNA_0 - \epsilon \frac{k_{-1}}{k_2};\quad \mathbb{L}_{2,2}^{2,1}(x_4) = 0 \\
\mathbb{U}_{2,1}^{2,1}(x_2) = \frac{M_0}{2} - x_4; \quad \mathbb{U}_{2,2}^{2,1} (x_2)= \frac{M_0}{2} - DNA_0 + \epsilon \frac{k_{-1}}{k_2}; \quad \mathbb{U}_{2,1}^{2,1}(x_4) = DNA_0; \quad \mathbb{U}_{2,2}^{2,1} (x_4) = \frac{M_0}{2} - x_2
\end{array}$

\end{example}



Using mass invariants to compute the most optimistic bound is done inductively over the $\mathcal{S}$ expressions of the candidate bounds, by applying usual formulae of interval arithmetics
 to propagate the $\dot{\min}$ and $\dot{max }$ operators. The resulting evaluation functions, which we call $f_{\dot\min}$ and $f_{\dot\max }$ respectively, are detailed in Appendix \ref{app:min-max}.





With all this in place, we can proceed to the definition of the reduced system.

\begin{definition}\label{def:red}\textbf{(Reduced system)}
Let $\mathcal{A}=(\mathcal{S},\mathbb{I},\mathbb{F},(\mathbb{E}_b))$ be a system of ordinary equations with equality constraints. The reduction of the system $\mathcal{A}$ is defined as the triple $(\mathcal{S}^\#,\mathbb{I}^\#,\mathbb{F}^\#)$, with:
\begin{enumerate}
\item $\mathcal{S}^\# = \{\underline{x_i}\mid x_i \in \mathcal{S}\}\cup\{\overline{x_i}\mid x_i\in \mathcal{S}\}$
\item $\mathbb{I}^\#:\mathcal{S}^\#\rightarrow\mathbb{R}^+$ is defined by $\mathbb{I}^\#(\underline{x_i})=\mathbb{I}^\#(\overline{x_i})=\mathbb{I}(x_i), \forall x_i \in\mathcal{S}$
h\item $\mathbb{F}^\#:\mathcal{S}^\# \rightarrow Expr_{\mathcal{S}^\#}$, defined as:
\begin{equation*}
\begin{cases}
\mathbb{F}^\#(\underline{x_i}) = f_{\dot\min} ([[\mathbb{F}_\downarrow^{1,1}(x_i)]_{\rho_1^\downarrow}]_{\rho_2^\downarrow},\ldots,[[\mathbb{F}_\downarrow^{p+1,n+1}(x_i)]_{\rho_1^\downarrow}]_{\rho_2^\downarrow})\\
\mathbb{F}^\#(\overline{x_i}) = f_{\dot\max } ([[\mathbb{F}_\uparrow^{1,1}(x_i)]_{\rho_1^\uparrow}]_{\rho_2^\uparrow},\ldots,[[\mathbb{F}_\uparrow^{p+1,n+1}(x_i)]_{\rho_1^\uparrow}]_{\rho_2^\uparrow})
\end{cases}
\end{equation*}

for every variable $x_i\in\mathcal{S}'$, where:
\begin{itemize}
\item $\rho_1^\downarrow = 
\begin{cases}
x_j \mapsto \dot{\max}(x_j,\dot{\max \limits_b}(\mathbb{L}_{i,b}^{k,l}(x_j))), &\text{ if } x_j \in t_{\downarrow,+}^{k,l,i} \\
x_j \mapsto \dot{\min}(x_j,\dot{\min\limits_b}(\mathbb{U}_{i,b}^{k,l}(x_j))), &\text{ if } x_j \in t_{\downarrow,-}^{k,l,i} 
\end{cases}$
\item $\rho_2^\downarrow = 
\begin{cases}
x_j \mapsto \underline{x_j}, &\text{ if } x_i=x_j \\
x_j \mapsto \underline{x_j}, &\text{ if } x_i\neq x_j, \text{ for positive polarity/sign occurences of } x_j \\
x_j \mapsto \overline{x_j}, &\text{ if } x_i\neq x_j,\text{ for negative polarity/sign occurences of } x_j
\end{cases}$
\item $\rho_1^\uparrow =
\begin{cases}
x_j \mapsto \dot{\min}(x_j,\dot{\min\limits_b}(\mathbb{U}_{i,b}^{k,l}(x_j))), &\text{ if } x_j \in t_{\uparrow,+}^{k,l,i} \\
x_j \mapsto \dot{\max }(x_j,\dot{\max\limits_b}(\mathbb{L}_{i,b}^{k,l}(x_j))), &\text{ if } x_j \in t_{\uparrow,-}^{k,l,i}
 \end{cases}$
\item $\rho_2^\uparrow = 
\begin{cases}
x_j \mapsto \overline{x_j}, &\text{ if } x_i=x_j \\
x_j \mapsto \overline{x_j}, &\text{ if } x_i\neq x_j, \text{ for positive polarity/sign occurences of } x_j \\
x_j \mapsto \underline{x_j}, &\text{ if } x_i\neq x_j, \text{ for negative polarity/sign occurences of } x_j
\end{cases}$
\end{itemize}
\end{enumerate}
\end{definition}

Intuitively, for each region $(k,l)$ of species $x_i$, the reduction method first replaces $\mathbb{F}(x_i)$  by the pair of tropicalized lower and upper bounds, $\mathbb{F}^{k,l}_\downarrow(x_i)$ and $\mathbb{F}^{k,l}_\uparrow(x_i)$, that result directly from the dominance inequalities that characterize the region. Then, $\mathbb{F}^{k,l}_\downarrow(x_i)$ and $\mathbb{F}^{k,l}_\uparrow(x_i)$ are refined, using the bounds on variables that can be deduced from the conservation laws of the original system. For example, replacing any occurence of a variable $x_j$ in $\mathbb{F}^{k,l}_\downarrow(x_i)$ with one of its expressions $\mathbb{E}_b(x_j)$ (or with its appropriate bound derived from $\mathbb{E}_b(x_j)$\footnote {$\mathbb{L}_{i,b}^{k,l}(x_j)$ for the positive occurences of $x_j$, and $\mathbb{U}_{i,b}^{k,l}(x_j)$ for its negative occurences}) results in another safe upper bound for $\mathbb{F}(x_i)$. By choosing the minimum such candidate bound, one obtains the most accurate, \emph{locally} safe upper bound. The same reasoning applies to the computation of lower bounds, but the $\min$ operation is replaced with $max $. 

In order to obtain safe (\emph{i.e.}, correct) global bounds, the least precise local bounds are chosen: the miminal lower, resp. the maximal upper bounds.

Finally, the interpretation of the variables is lifted over the hyper-faces. Any occurence of $x_i$ is replaced with its analogue corresponding to the hyperface we want to bound, while the others are replaced to their analogue given by the polarity of their occurence, as explained in Note \ref{note1}. 


\begin{theorem} Let $\mathcal{A}=(\mathcal{S},\mathbb{I},\mathbb{F},(\mathbb{E}_b))$ be a system of ordinary equations with equality constraints. Let $(\mathcal{S}^\#,\mathbb{I}^\#,\mathbb{F}^\#)$ be a reduction of the system $\mathcal{A}$. 

Let $f$ be a function from the set $\mathcal{S}$ into the set $\mathbb{R}^{\mathbb{R}^+}$ s.t. for every variable $x_i \in \mathcal{S}$, we have:
\begin{equation*}
\begin{cases}
f(x_i)(0) = \mathbb{I}(x_i) \\
\frac{df(x_i)}{dt}(t) = \mathbb{F}[x_i\mapsto f(x_i)(t)]
\end{cases}
\end{equation*}

and $f\#$ be a function from the set $\mathcal{S}^\#$ into the set $\mathbb{R}^{\mathbb{R}^+}$ s.t. for very abstract variable $x_i\#\in\mathcal{S}^\#$, we have: 

\begin{equation*}
\begin{cases}
f(x_i^\#)(0) = \mathbb{I^\#}(x_i^\#) \\
\frac{df^\#(x_i^\#)}{dt}(t) = \mathbb{F}^\#[x_i^\#\mapsto f^\#(x_i^\#)(t)]
\end{cases}
\end{equation*}

Under these assumptions, we have that for every variable $x_i\in\mathcal{S}$ and every time $t\in\mathbb{R}^+$:
\begin{equation*}
f^\#(\underline{x_i})(t))\leq f(x_i)(t) \leq f^\#(\overline{x_i})(t), 
\end{equation*}
\emph{i.e.}, the reduced system provides sound lower and upper bounds for the concentration of the original system's species.
\end{theorem}

\begin{example}\label{ex:DNAmodel} We apply our method on the DNA example constructed in Sect.\ref{sub:DNA}, for different values of the scale separating constant $\epsilon$, and for arbitrarily chosen reaction rate constants $k_1=k_2=k_3=0.1, k_{-1}=0.01, k_{-2}=0.00001$ and initial concentrations $[M]_0=1, [DNA]_0=0.05$. We show in Fig.\ref{fig:DNA} the time evolutions on the bounds on the concentration of the product $[P]$, \emph{i.e.} the variable $x_5$. We notice once again that the results become more precise as $\epsilon$ decreases, \emph{i.e.} as the monomials become more separated. As an example, the ODE of the lower bound of species $x_2$ can be found in Appendix \ref{app:lowerx2}.

\begin{figure}\centering
 \includegraphics[width=0.7\columnwidth, height=6cm]{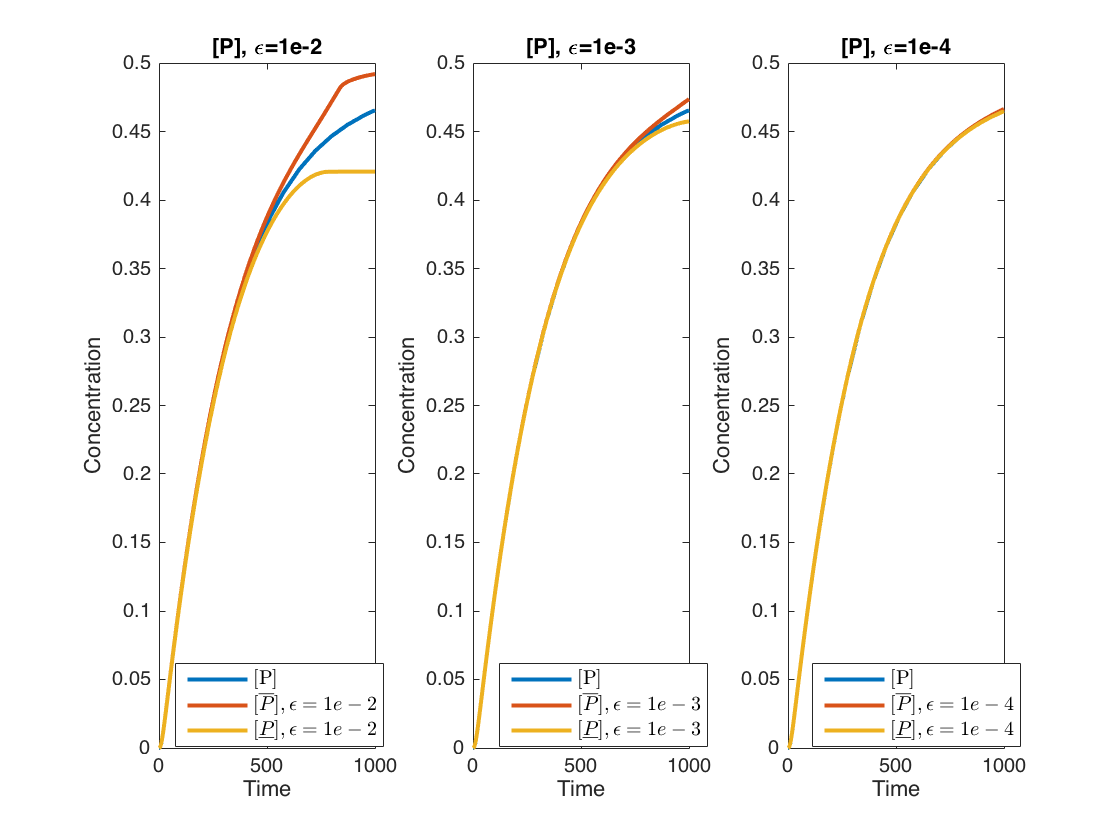}
 \caption{Bounds on the concentration of P, in the DNA example, obtained by simulating the ODE of tropicalized bounds for different values of the scale separation constant $\epsilon$, rate constants $k_1=k_2=k_3=0.1, k_{-1}=0.01, k_{-2}=0.00001$ and initial concentrations $[M]_0=1, [DNA]_0=0.05$.}  
\label{fig:DNA}
\end{figure}

\end{example}

\section{Error estimates of tropicalized systems using conservative numerical approximations}

Our approach also serves as a heuristics for quantifying errors of tropicalization approaches for biochemical model reduction, provided a slight modification is applied to Def.\ref{def:refbounds}. Instead of computing bounds using only variables from the support of the tropicalized bounds, one can use the equality constraints ${\mathbb\{E_b\}}$, to refine the accuracy of bounds. The resulting model presents a trade-off: it introduces new variables w.r.t. the support of the tropicalized bounds, albeit exclusively in the form of conservation laws which are always linear functions, but gains in bound accuracy. As such, the approximation error/accuracy of a given reduction method can be assessed by checking if the reduced trajectory lies between the lower and upper bounds computed by our method. 

\begin{example}\label{ref:MMerror}
It is well known that the Michaelis-Menten reduction is valid only under the QSS and QE assumptions. In Fig.\ref{fig:MMerror}, we simulate the reduced Michaelis-Menten system (\ref{sys:MM-QSS}), as well as our modified reduced system, as presented above, for a set of initial conditions that no longer satisfy the QSS assumption, \emph{i.e.} the total enzyme concentration is comparable to the total substrate concentration. As expected, the reduced Michaelis-Menten system no longer represents a good approximation of the initial enzymatic system (\ref{sys:orig}); this is reflected by the fact that the trajectory of the reduced model does not lie between the lower and upper bounds computed by our approach.
\end{example}

\begin{figure}[!ht]\centering
 \begin{subfigure}{.5\textwidth}
  \centering
  \includegraphics[width=\linewidth]{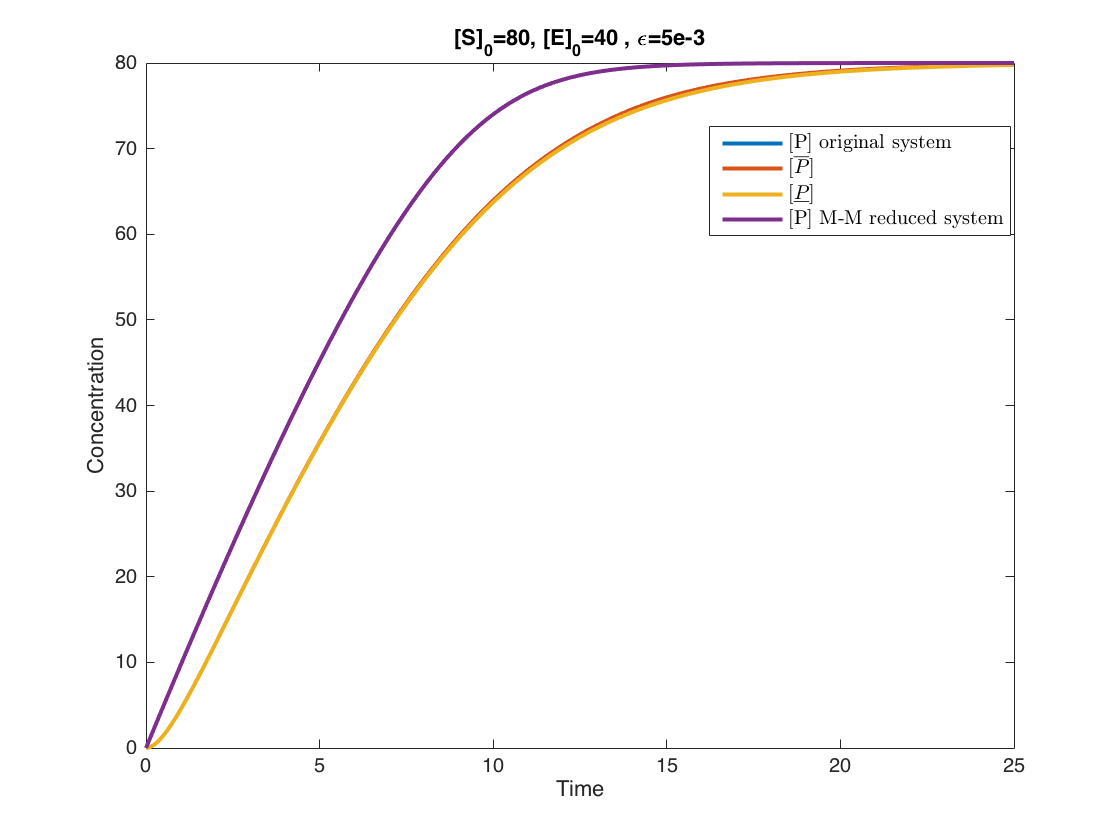}
  \label{fig:MMerror}
\end{subfigure}%
\begin{subfigure}{.5\textwidth}
  \centering
  \includegraphics[width=\linewidth]{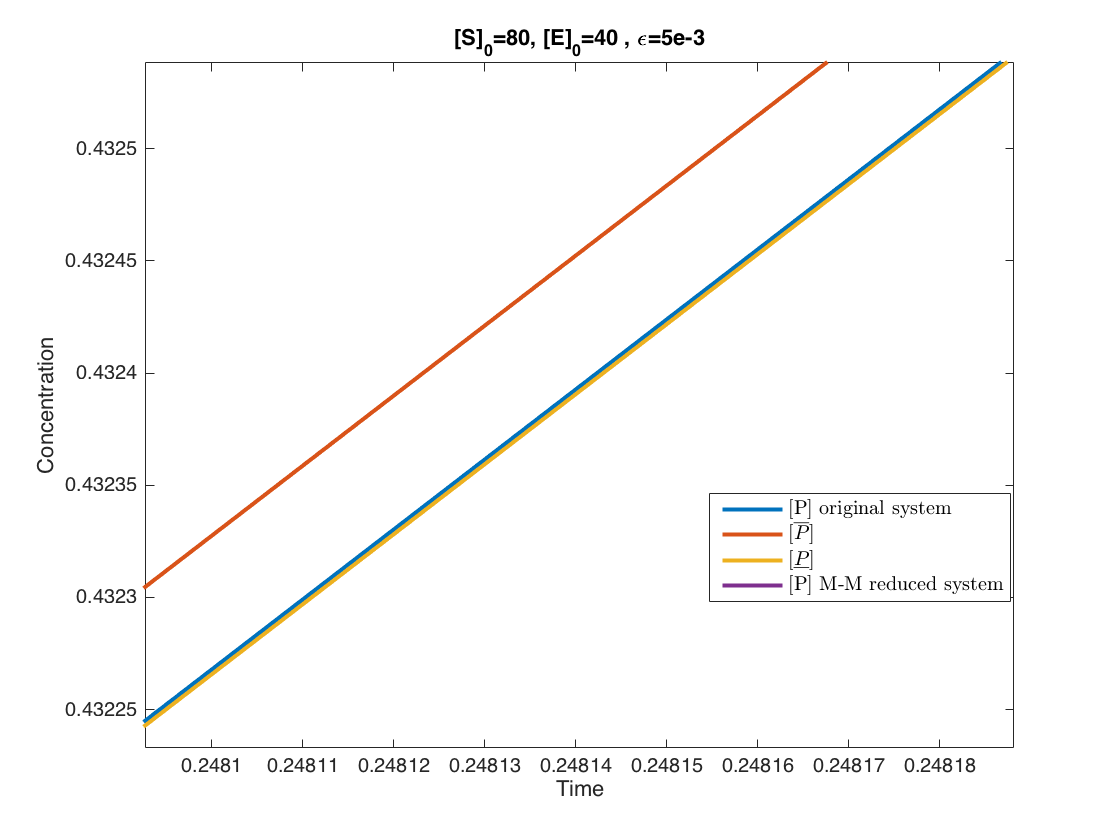}
  \label{fig:MMerror-zoom}
\end{subfigure}
 \caption{Estimating the accuracy of the Michaelis-Menten approximation: bounds on the concentration of [P], with respect to simulation time, for $\epsilon=5\cdot10^{-4}$, rate constants $k_1=0.017, k_{-1}=0.0017, k_2=0.3$, and initial concentrations that do not satisfy the QSS condition: $[S]=80,[E]=40,[E:S]=0,[P]=0$. (left) Whereas the original system's trajectory lies between the lower and upper bound given by our method, this is not the case for the classical Michaelis-Menten approximation. Thus, as expected, one can conclude that if the QSS condition is not met, the Michaelis-Menten approximation is inaccurate. (right) Zoomed in version, showing the enclosed original trajectory (in blue)}  
\label{fig:MMerror}
\end{figure}

\subsection{Tyson's Cell Cycle Model}

The tropicalization heuristics can be difficult to justify by rigourous estimates, although this is possible in some cases\cite{vnoel}. For example, the existence of tropical varieties - the set of points $x\in\mathbb{R}^n$ where at least two monomials of $P^{-/+}$ are equal- can lead to sliding modes, which in turn represent challenges in providing accuracy justifications for hybrid models obtained using tropical ideas. Sliding modes are well known phenomena in ODEs with discontinuous vector fields, in which the dynamics can follow discontinuity hyper-surfaces where the vector field is not defined; what's more, the conditions for the existence of sliding modes are usually intricate. As noted in \cite{vnoel2}, sliding modes can have a nefarious effect on the behavior of the tropicalized system: tropical varieties (\emph{i.e.} tropical curves) decompose the state space into sectors corresponding to the smooth modes of the hybrid tropicalized system, which passes from one type of smooth dynamics to another intrinsically, when the trajectory attains the tropical curve. However, if certain conditions w.r.t. the sliding modes are fulfilled, the trajectory can continue along some tropical curve instead of changing sector, which further deviates the reduced system's trajectory from the original one (see Figure 1 in \cite{vnoel2}, for an example). 

In \cite{vnoel2}, such phenomena become apparent when tropicalization is applied  to the minimal cell cycle model proposed by Tyson\cite{Tyson}, in order to obtain a reduced hybrid model. The Tyson model describes the interplay between cyclin and cyclin dependent kinase cdc2 during the progression of the cell cycle, and demonstrates the existence of three possible regimes, that can be associated to different phases in the cell life: the biochemical system can either  function as an oscillator, converge to a steady state, or behave as an excitable switch. The three possible behaviours  can be associated to early embryos rapid division, arrest of unfertilised eggs and growth controlled division of somatic cells, respectively. 

The dynamics of this non-linear model with rational reaction rates contains 6 species and 9 reactions, and is described by the following system of polynomial differential equations: 

\begin{equation}\label{eq:Tyson}
\begin{cases}\frac{dy_1}{dt} = k_6\cdot y_4 - k_8\cdot y_1 + k_9\cdot y_2\\
\frac{dy_2}{dt}= -k_3\cdot y_2\cdot y_5 + k_8\cdot y_1 - k_9\cdot y_2\\
\frac{dy_3}{dt}= k_3\cdot y_2\cdot y_5 - k’_4\cdot y_3 - k_4\cdot y_4^2\cdot y_3 \\
\frac{dy_4}{dt}= k’_4\cdot y_3 + k_4\cdot y_4^2\cdot y_3 - k_6\cdot y_4\\
\frac{dy_5}{dt} = k_1 -k_3\cdot y_2\cdot y_5\\
\frac{dy_6}{dt}=  k_6\cdot y_4 - k_7\cdot y_6
\end{cases},
\end{equation}

and has the conservation law $y_1(t)+y_2(t)+y_3(t)+y_4(t)=1$, where the value 1 denoting the total initial concentration of kinase cdc2 (\emph{i.e.} $y_1(0)+y_2(0)+y_3(0)+y_4(0)$)  was chosen by convenience. The values of the reaction rates constants are fixed as to have the model display the oscillatory behavior: $k_1=0.015, k_3=200, k_4=180,  k'_4=0.018,  k_6=1, k_8=10^3, k_9=10^6$.

\begin{figure}[!ht]\centering
 \includegraphics[width=0.6\columnwidth, height=5cm]{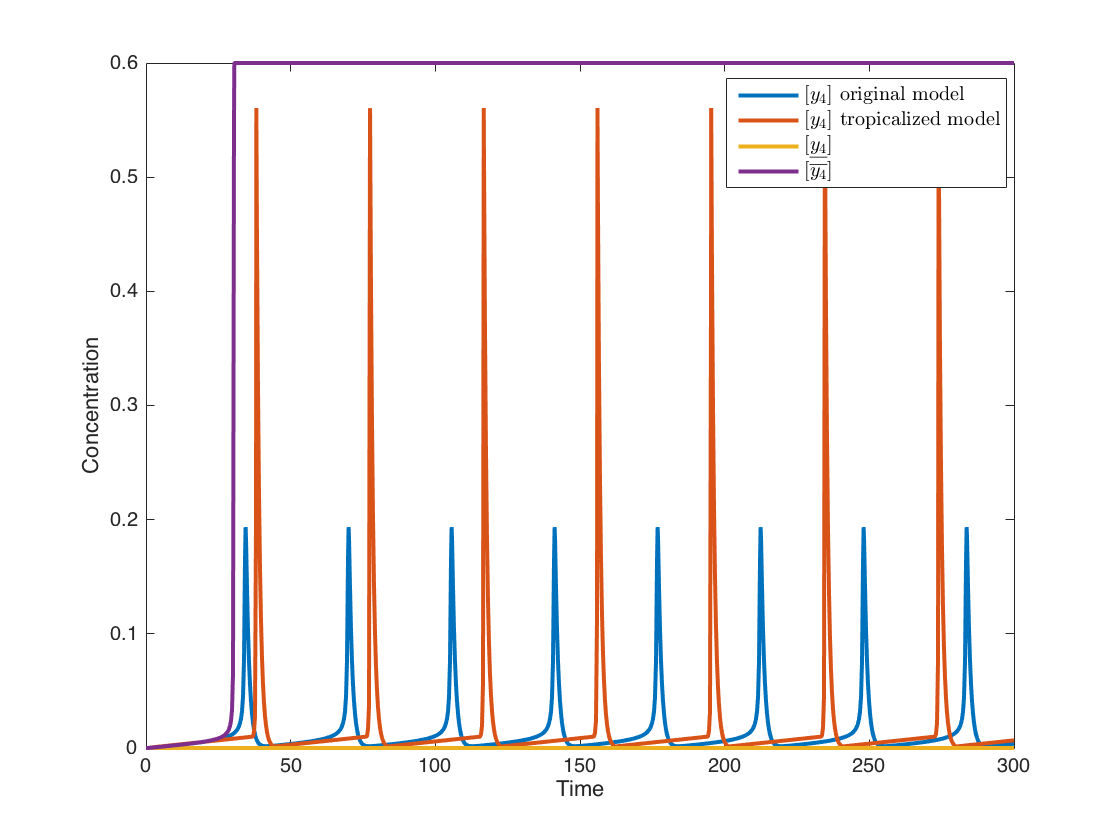}
 \caption{Estimating the accuracy of the tropicalized Tyson model: bounds on the concentration of $[y_4]$, with respect to simulation time, for $\epsilon=10^{-3}$.}  
\label{fig:Tysonerror}
\end{figure}

In \cite{vnoel,vnoel2}, a hybrid model of the Tyson cell cycle is obtained by detecting and eliminating QSS species of the original model, pruning dominated monomials, and then ultimately tropicalizing the reduced-size model. Besides having the inconvenient of analyzing trajectories of the original model in order to detect QSS species, the reduced model suffers from the the sliding mode-related issues mentioned above: although both the smooth (original)  and the reduced system exhibit oscillating behavior and have stable periodic trajectory (\emph{i.e.} limit cycle), the period of the tropicalized limit cycle is different with respect to that of the smooth cycle, due to the tropicalized trajectory sliding along the tropical manifolds instead of changing sectors.  Having different oscillation periods means in turn that assessing the accuracy of the tropicalized reduced model is not a trivial question, as the distance between original and tropicalized trajectories is variable from cycle to cycle (as can be seen in Figure\ref{fig:Tysonerror}). What's more, it can also provide an indication of the poor performance of tropicalization based reduction methods when dealing with more complex systems, such oscillating systems.

Indeed, by applying our method to the original Tyson model, we are able to effectively provide guarantees on the reduced model, albeit not very strong ones: this could be interpreted as an indication of the poor accuracy of the tropicalized Tyson model.
In Figure \ref{fig:Tysonerror}, we plot the bounds for the concentration of species $y_4$ obtained using our method, in order to compare the trajectory of the original model to the one of the hybrid one that can be found in \cite{vnoel}. The lack of oscillating behavior in the computed bounds could intuitively be explained by the difference in period of the original and reduced systems, that causes a shift at every cycle in the tropicalized trajectory w.r.t. the original behavior. 
Nonetheless, the obtained bounds accurately capture the amplitude of the tropicalized model. One also notes that the time points where the upper bound, respectively the tropicalized system, begin to diverge w.r.t. the original trajectory coincide. 

From a more practical point of view, we note that while simulation of the tropicalized Tyson model proposed in \cite{vnoel2} was performed in 354.876706 seconds, on a 2.2 GHz Intel Core i7 processor and 16 GB 1600 MHz DDR3 memory, simulation of the model obtained via our method was performed in only 9.775511 seconds using the same numerical integration method (\emph{i.e.} Matlab's \emph{ode15s}), thus providing a significant improvement in computation time.

We note that an alternative reduced model is obtained in \cite{vnoel}, using tropical equilibration, that circumvents the need to simulate the original system. We plan to include tropical equilibration techniques in future work.

\section{Comparison with existing methods}
We mentioned previously that numerical errors stemming from numerical integration are ignored herein. Indeed, numerical integration methods, while heavily used, only provide \emph{approximations} of the solution of the initial value problem (IVP) of ODE systems. Even when using variable-step size methods, there are no guarantees that the approximate solution computed by the chosen method is close to the actual solution.   In order to solve the drawbacks associated to traditional ODE solvers/numerical solutions of IVP, interval numerical methods for IVP are used for computing validated enclosures of the solution of an IVP for an ODE. For example, the VNODE-LP\cite{vnode-lp} C++ solver proves that a unique solution to a problem exists, and then computes rigorous bounds that are guaranteed to contain it. Such bounds can then be used to help prove theoretical results, check if a solution satisfied a condition in a safety-critical calculation, or simply to verify the results produced by a traditional ODE solver. Another example of such software is the CAPD library \cite{capd}. Both represent well-established software for computing enclosures of generic ODE systems, and are integrated in various SMT solvers (e.g., iSAT\cite{isat}, dReal\cite{dreal}). For a more comprehensive state of the art on such methods, the reader is refered to \cite{ned}.

Interval methods for IVPs for ordinary differential equations are typically based on Taylor series expansions, which require the computation of Taylor coefficients up to some order $k$. Given a final time point, the aim is to compute interval vectors that are guaranteed to contain the solution to a given IVP, at all intermediary points. In order to compute such interval vectors, interval propagation methods are used to enclose roundoff and truncation errors in the computed bounds, and thus obtain rigorous bounds on the true solution of the ODE.


In our approach, instead of interpreting the differential equations over the state of the system, the interpretation is lifted conservatively over each hyper-face of the hyper-box  abstracting the system state (\emph{i.e.}, we over-approximate the derivatives only on the hyper-faces). When compared to our method, interval propagation methods over-approximate the partial derivatives of the function over the whole enclosing hyper-box, instead of doing so only on the hyper-faces. This in turn means that our approach computes tighter bounds than those computed by interval methods for IVPs.  

\begin{figure}
        \centering
        \begin{subfigure}[b]{0.475\textwidth}
            \centering
            \includegraphics[width=\textwidth]{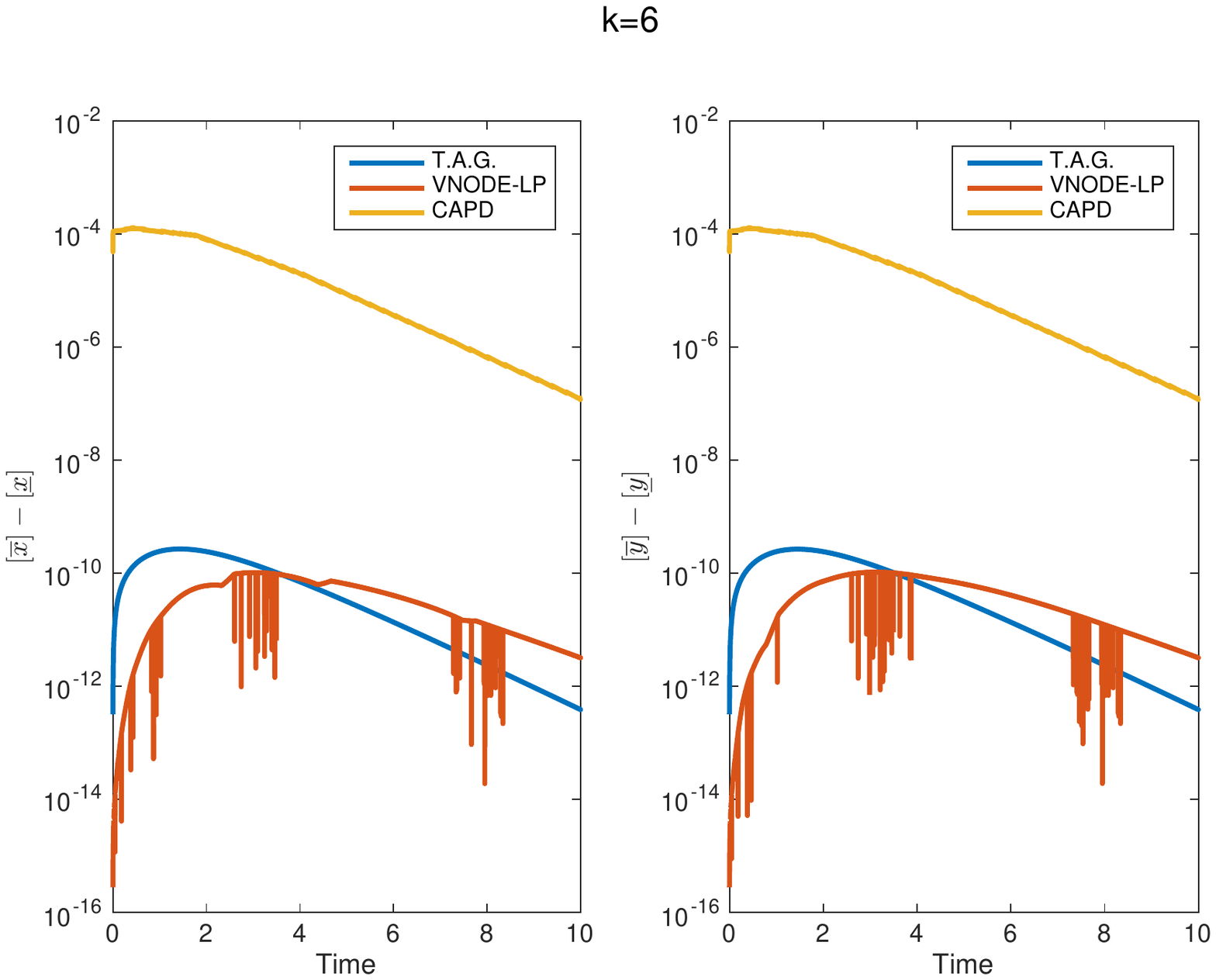}   
        \end{subfigure}
        \hfill
        \begin{subfigure}[b]{0.475\textwidth}  
            \centering 
            \includegraphics[width=\textwidth]{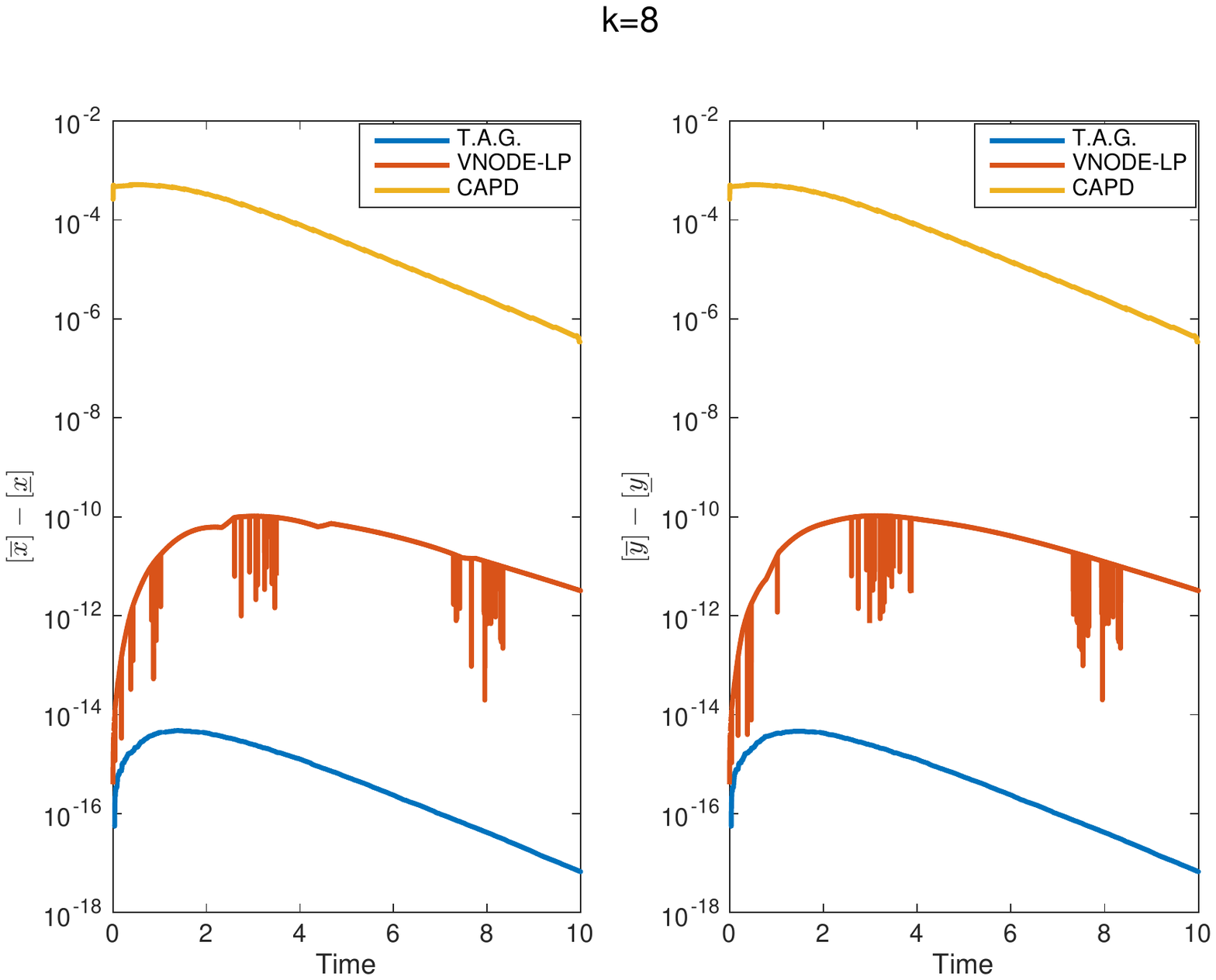}
        \end{subfigure}
        \vskip\baselineskip
        \begin{subfigure}[b]{0.475\textwidth}   
            \centering 
            \includegraphics[width=\textwidth]{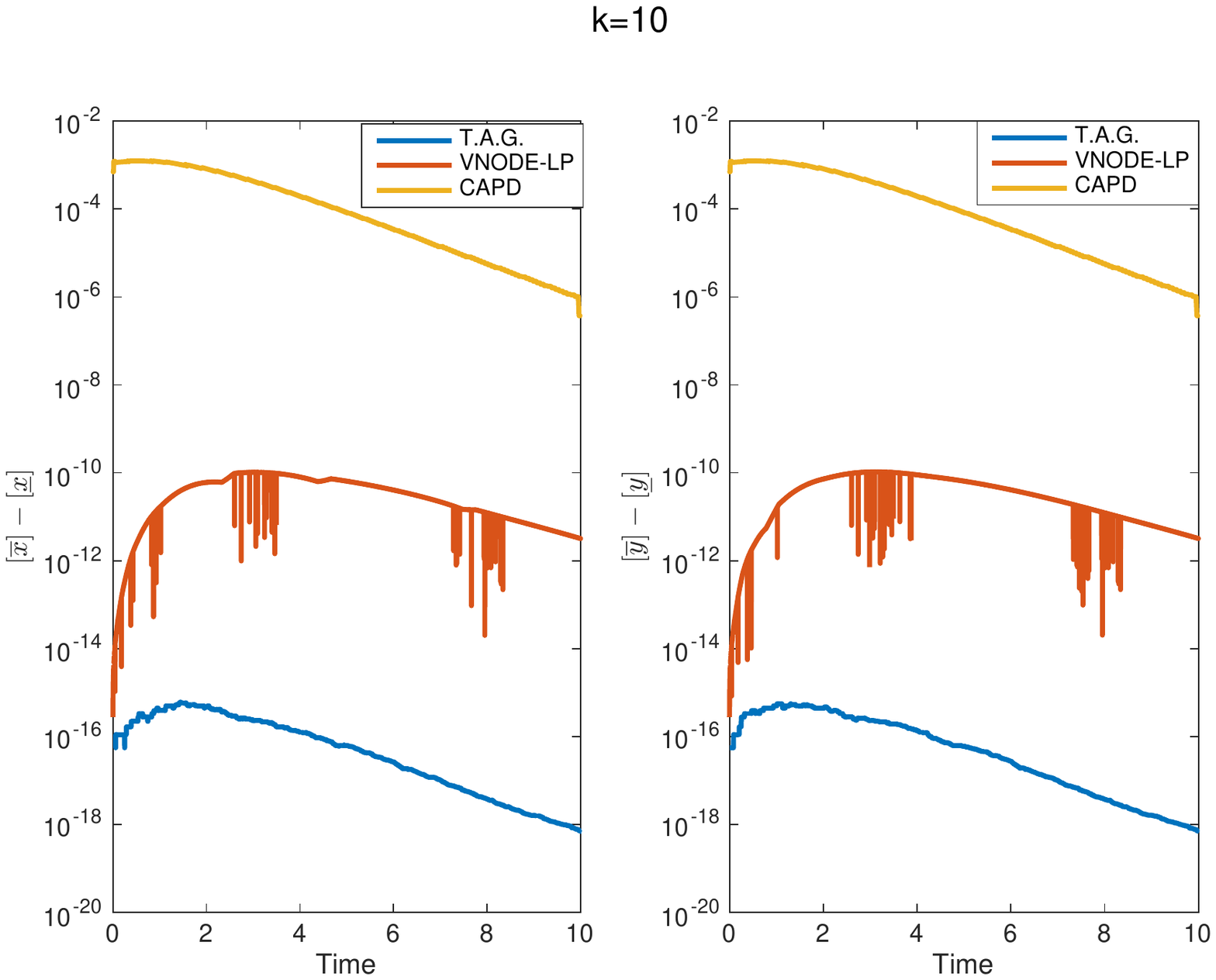}
        \end{subfigure}
        \quad
        \begin{subfigure}[b]{0.475\textwidth}   
            \centering 
            \includegraphics[width=\textwidth]{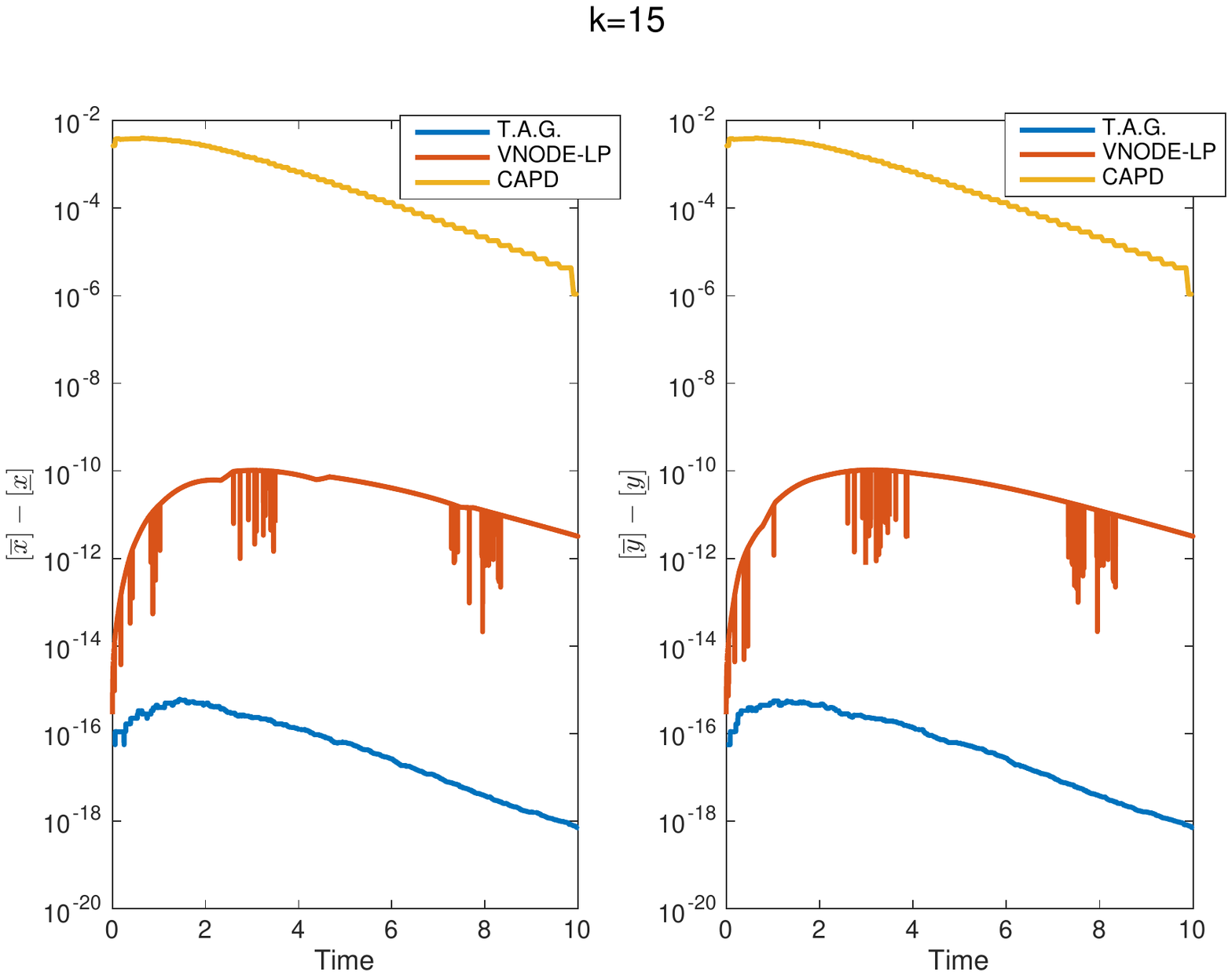}
        \end{subfigure}
        \vskip\baselineskip
        \centering            
        \includegraphics[width=0.5\textwidth]{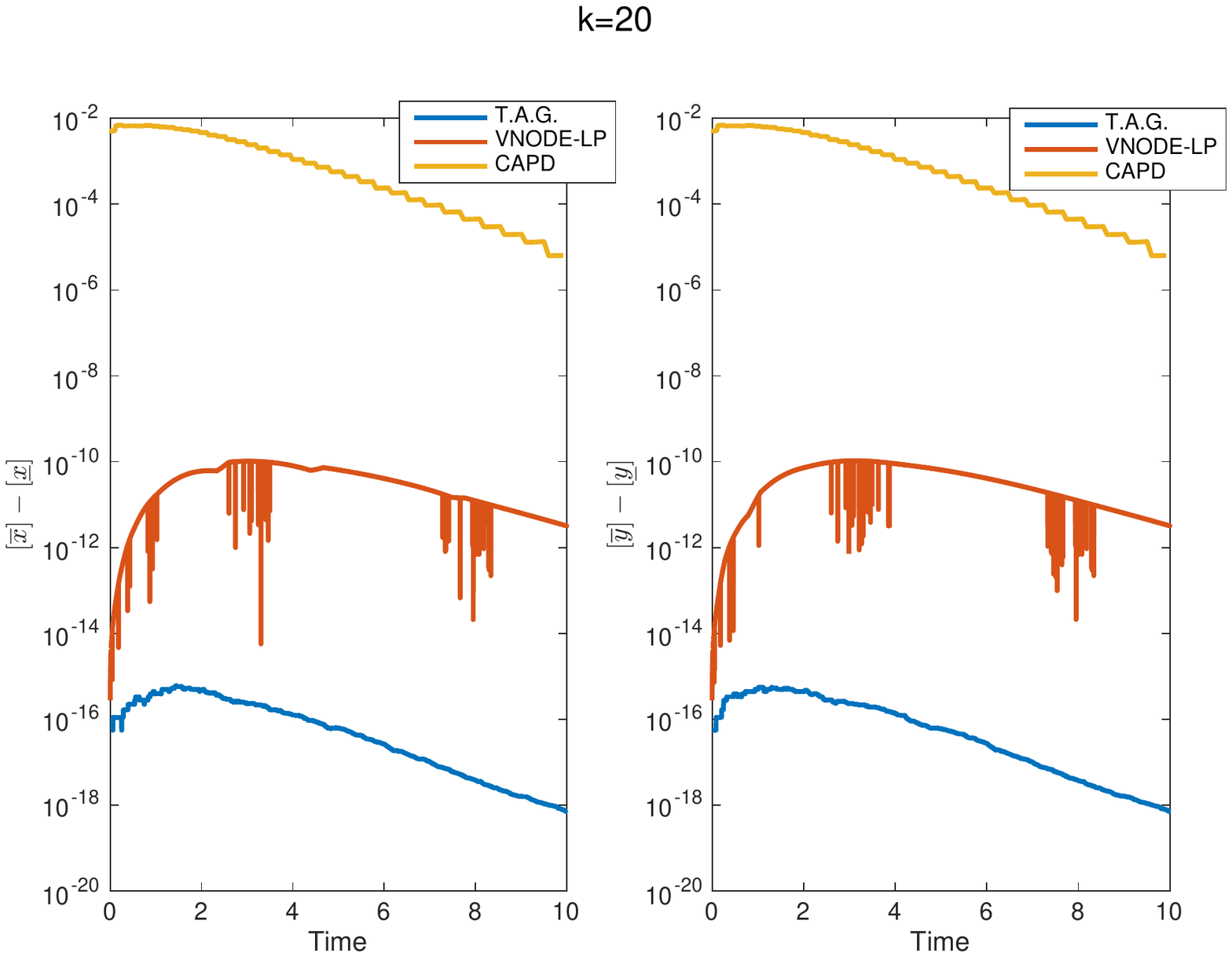}   

    \caption{Comparison of our method - which we denote by "T.A.G."- with existing ODE enclosure methods CAPD and VNODE-LP. For different values of the Taylor expansion order $k$, used to derive bounds on the system in Example 5.1, our method ultimately out-performs both CAPD and VNODE-LP, in terms of accuracy; for $k=6$, our method is initially outperformed by VNODE-LP on the studied example, but ultimately proves to be more efficient, for the time frame $t>4$. The exponent on the y-axis is an indicator of the decimal precision of the methods (i.e. a value of $10^{-15}$ means that the first 15 decimals of the computed lower and upper bounds are identical). We also note that both CAPD and VNODE-LP only allow values of $k>4$.} 
        \label{fig:compare}
    \end{figure}

We demonstrate our claim with the following example:
\begin{example}
Let us consider the following initial value problem :

\begin{center}
$\begin{cases}
\frac{dx}{dt} =  y\cdot(2-cos(y)) - x\cdot(2-sin(y))\\
\frac{dy}{dt} =  x\cdot(2−cos(y)) − y\cdot(2−sin(y)) \\
x(0)=y(0)=1
\end{cases}$
\end{center}

As presented in Section 3, our framework can be decomposed in two independents parts: the first part consists in synthesizing bounds on the derivatives of the original system, and the second part deals with the propagation of said bounds, in order to obtain the enclosing system. 

As our goal is to better understand and evaluate tropicalization approaches for biochemical model reduction, so far we chose to focus on bounds obtained by using dominance relations between monomials. The second part of our method simply represents an improved alternative to existing ODE enclosure methods, as explained above, and as such can be used in such methods in order to get better enclosure results. 

For example, in order to compare the performance of our method to that of VNODE-LP and CAPD, instead of using dominance relations to derive inequality constraints on species' concentrations, we now use the Taylor Series expansion with $k$ terms ($k$ will serve as a parameter) for the functions $sin$ and $cos$, in order to derive bounds on $\frac{dx}{dt}$ and $\frac{dy}{dt}$:

\begin{center}
$\begin{cases}
sin(x)\approx x-\frac{x^3}{3!}+\frac{x^5}{5!}-\frac{x^7}{7!}+\frac{x^9}{9!}-\ldots=\sum\limits_{n=0}^{k-1}(-1)^n\frac{x^{2n+1}}{(2n+1)!}\\
cos(x)\approx 1-\frac{x^2}{2!}+\frac{x^4}{4!}-\frac{x^6}{6!}+\frac{x^8}{8!}-\ldots=\sum\limits_{n=0}^{k-1}(-1)^n\frac{x^{2n}}{(2n)!}
\end{cases}$
\end{center}

Then, for a fixed order $k$ and an 
$\epsilon$, instead of using dominance-related inequalities with our method, one can use the following inequalities:
\newline

\begin{center}
\begin{align*}
\sum\limits_{n=0}^{k-1}(-1)^n\frac{x^{2n+1}}{(2n+1)!}-\epsilon &\leq sin(x)\leq\sum\limits_{n=0}^{k-1}(-1)^n\frac{x^{2n+1}}{(2n+1)!}+\epsilon\\
\sum\limits_{n=0}^{k-1}(-1)^n\frac{x^{2n}}{(2n)!}-\epsilon &\leq cos(x)\leq\sum\limits_{n=0}^{k(-1}-1)^n\frac{x^{2n}}{(2n)!}+\epsilon
\end{align*},
\end{center}

where $\epsilon = (-1)^{2k+1}\frac{x^{2k+1}}{(2k+1)!}cos(c_x)$ for $c_x\in[0,x]$ is the residual for the Taylor expansion, and  can be bound by $\epsilon\leq\frac{|x|^{2k+1}}{(2k+1)!}$, which in turn is $\leq (2k+1)!^{-1}$ for $x\in[-1,1]$ \cite{}. 

Our method then proceeds as usual to the computation of ODEs for bounds on the concentrations of $x$ and $y$.

In Fig.\ref{fig:compare}, we compare the accuracy of our method to that of VNODE-LP and CAPD, for different values of the order $k$. The accuracy is given by the tightness of bounds, which can be evaluated by computing the difference between the upper and  lower bounds, during a simulation. The results indicate that, when compared to existing enclosure interval methods,  our approach represents a consistent improvement of several orders of magnitude, across different values of $k$.
\end{example}

\section{Conclusion and outlook}

In this paper, we present an approximation method for biochemical networks,  which can also serve as a technique for evaluating the faithfulness of existing tropicalization reduction methods that do not involve guarantees.
Our approach relies on the multiscaleness property of biochemical systems. Tropical geometry offers a natural framework for studying such networks. Tropical approaches \cite{vnoel3,rad} can guide model reduction of ODE systems, by using time- and concentration scales separation to identify and neglect equation terms whose values are significantly smaller than those of other terms of the same equation. This leads to partitioning the state space into different regions, according to which term dominates the others. A similar approach is employed in our method, but instead of neglecting the dominated terms, we propose to conservatively bound their value using an amortizing scale separation constant and the value of the dominant terms. These bounds can be further refined by incorporating the conservation laws of the initial system. The resulting approximated model is composed of two-term ODEs (which we call tropicalized), which by construction provide time-dependent lower and upper bounds for the concentration of the initial system's species. As such, our approach can also serve to test the accuracy of other given reduction methods, while circumventing the execution of the original system: the suitability of a reduction will be confirmed if the reduced model's trajectory lies between the bounds provided by our abstraction. 

We have tested our approach on the classical Michaelis-Menten system, a simple extension of it, and Tyson's cell cycle model. Our method can be easily automatized, either using static analysis, or existing symbolic math tools\footnote{for example, Matlab's Symbolic Math Toolbox} ; as such, Definitions 3.1-3.11  are written in an operational-semantics style, as to describe the different procedures composing the algorithm that implements our method. A tool that automatizes our approach is currently being developed. Further work also includes expanding the case studies to larger networks, possibly with no conservation laws.


%
%
%

\begin{thebibliography}{15}\label{bibliography}












\bibitem{max-plus}F. Baccelli, G. Cohen, G. J. Olsder, J.-P. Quadrat. \emph{Synchronisation and linearity}. Wiley, 1992

\bibitem{beica}A. Beica, C. C. Guet, T. Petrov. \emph{Efficient reduction of kappa models by static inspection of the rule-set}. In: A. Abate, D. Safranek (Eds.), Hybrid Systems Biology - Fourth International Workshop, HSB 2015, Madrid, Spain, September 4-5, 2015. Revised Selected Papers, Vol. 9271 of Lecture Notes in Computer Science, Springer, 2015, pp. 173--191

\bibitem{bngl} M.L. Blinov, J.R. Faeder, B. Goldstein, and W.S. Hlavacek. \emph{Bionetgen: software for rule-based modeling of signal transduction based on the interactions of molecular domains}. Bioinformatics, 20(17), 2004.

\bibitem{boseung} C. Boseung, GA Rempala, JK Kim. \emph{Beyond the Michaelis-Menten equation: Accurate and efficient estimation of enzyme kinetic parameters}. Scientific Reports, 7(1), 2017.


\bibitem{qss} G.E. Briggs, J.B.S. Haldane.\emph{A note on the kinematics of enzyme action}. Biochem J. 19 (2): 338--339, 1925. 

\bibitem{capd} Computer Assisted Proofs in Dynamics (CAPD) Library. \href{url:}{http://capd.ii.uj.edu.pl}

\bibitem{ken} K. Chanseau Saint-Germain, J. Feret, \emph{Conservative numerical approximations of the differential semantics in biological rule-based models}, 2016

\bibitem{chaouiya} C. Chaouiya. \emph{Petri net modelling of biological networks}. Briefings in Bioinformatics, Volume 8, Issue 4, 1 July 2007, Pages 210--219.

\bibitem{kappa3} V. Danos, J. Feret, W. Fontana, and J. Krivine. \emph{Scalable simulation of cellular signaling networks}, invited paper. In Proc. APLAS--07, volume 4807 of LNCS. Springer, 2007.

\bibitem{kappa1}V.Danos, J. Feret, W. Fontana, R. Harmer, and J. Krivine. \emph{Rule based modeling of biological signaling}. In Proc. CONCUR'07, volume 4703 of LNCS, 2007.

\bibitem{kappa2} V. Danos and C. Laneve. \emph{Formal molecular biology}. TCS, 325(1), 2004

\bibitem{isat}M. Franzle, C. Herde, T. Teige, S. Ratschan, and T. Schubert. \emph{Efficient solving of large non-linear arithmetic constraint systems with complex boolean structure}. In Journal on Satisfiability, Boolean Modeling, and Computation, 1(3-4):209–236, 2007.

\bibitem{feret2009internal} J. Feret, V. Danos, J. Krivine, Jean, R. Harmer, W. Fontana. \emph{Internal coarse-graining of molecular systems}. PNAS, 2009

\bibitem{feret2012lumpability} J. Feret, T. Henzinger, H. Koeppl, Heinz, T. Petrov. \emph{Lumpability abstractions of rule-based systems}. Elsevier, 2012

\bibitem{dreal}S. Gao, S. Kong, E.M. Clarke.\emph{dReal: An SMT Solver for Nonlinear Theories over the Reals}. In: Bonacina M.P. (eds) Automated Deduction – CADE-24. CADE 2013. Lecture Notes in Computer Science, vol 7898. Springer, Berlin, Heidelberg

\bibitem{green} D.G. Green. \emph{Cellular automata models in biology}. Mathematical and Computer Modelling,Volume 13, Issue 6,
1990, pp. 69--74

\bibitem{hlavacek2003complexity} W.S.Hlavacek, J.R. Faeder, M.L.Blinov, A.S. Perelson, B.Goldstein, \emph{The complexity of complexes in signal transduction}. Wiley, 2003

\bibitem{qe} J. Keener, J. Sneyd.\emph{Mathematical Physiology: I: Cellular Physiology} (2 ed.). Springer, 1994.

\bibitem{kirk}M. Kirkilionis, S. Walcher. \emph{On comparison systems for ordinary differential equations}. Journal of Mathematical Analysis and Applications, 299 (1), 2004, pp.157 --173. 

\bibitem{kurtz1970solutions} T.G.Kurtz. \emph{Solutions of ordinary differential equations as limits of pure jump Markov processes}. Cambridge University Press, 1970

\bibitem{lit} G.L. Litvinov. \emph{Tropical Mathematics, Idempotent Analysis, Classical Mechanics and Geometry}. E-print arXiv:1005.1247, 2010.

\bibitem{matlab} \emph{MATLAB and Statistics Toolbox Release 2015b}, The MathWorks, Inc., Natick, Massachusetts, United States, URL: \href{https://fr.mathworks.com/help/matlab/ref/ode15s.html}{\texttt{https://fr.mathworks.com/help/matlab/ref/ode15s.html}}

\bibitem{mm} L. Michaelis, M.L. Menten. \emph{Die Kinetik der Invertinwirkung}. Biochem Z. 49: 333--369, 1913.

\bibitem{vnode-lp}N.S. Nedialkov.\emph{VNODE-LP: A Validated Solver for Initial Value Problems in Ordinary Differential Equations}. Technical Report CAS-06-06-NN, 2006.

\bibitem{ned}N. S. Nedialkov. \emph{Interval Tools for ODEs and DAEs}. In: 12th GAMM - IMACS International Symposium on Scientific Computing, Computer Arithmetic and Validated Numerics (SCAN 2006), Duisburg, 2006.

\bibitem{vnoel}V. No\"{e}l. \emph{Mod\`{e}les r\'{e}duits et hybrides de r\'{e}seaux de r\'{e}actions biochimiques : applications \`{a} la mod\'{e}lisation du cycle cellulaire.} Bio-Informatique, Biologie Syst\`{e}mique. Universit\'{e} Rennes 1, 2012. (PhD Thesis)

\bibitem{vnoel2} V. No\"{e}l, D. Grigoriev, S. Vakulenko, O. Radulescu. \emph{Tropical geometries and dynamics of biochemical networks application to hybrid cell cycle models}. Electronic Notes in Theoretical Computer Science, 2012.

\bibitem{vnoel3} O. Radulescu, A. N. Gorban, A. Zinovyev, V. Noel. \emph{Reduction of dynamical biochemical reactions networks in computational biology}, Frontiers in Genetics 3 (131) (2012) 17.

\bibitem{rad} O. Radulescu, S. Vakulenko, D. Grigoriev. \emph{ Model reduction of biochemical reactions networks by tropical analysis methods}. Mathematical Modelling of Natural Phenomena 10 (3), 2015.


\bibitem{wang} Rui-Sheng Wang et al. \emph{Boolean modeling in systems biology: an overview of methodology and applications}. 2012 Phys. Biol. 9 055001

\bibitem{Tyson} J.J. Tyson. \emph{Modeling the cell division cycle: cdc2 and cyclin interactions}. Proceedings of the National Academy of Sciences of the United States of America, 88(16):7328, 1991.

\bibitem{wyn} ML Wynn, N Consul, SD Merajver, S Schnell. \emph{Logic-based models in systems biology: a predictive and parameter-free network analysis method}. Integrative biology: quantitative biosciences from nano to macro. 2012; 4(11):10.1039/c2ib20193c. 

\end{thebibliography}
%

\begin{appendices}
\section{Symbolic propagation of min and max }\label{app:min-max}
The propagation of the $\min$ and $max $ operations over $\mathcal{S}$-expression is done using the evaluation functions $f_{\dot\min}$ and $f_{\dot\max }$, which are defined by mutual induction over the syntax of $\mathcal{S}$-expressions denoting monomials:

\begin{enumerate}
\item $\forall e_1, e_2, \ldots, e_k \in Expr_{\mathcal{S}}$,
	\begin{itemize}
	\item $f_{\dot\min}(e_1,e_2,\ldots,e_k)\triangleq\dot\min(e_1,e_2,\ldots,e_k)$
	\item $f_{\dot\max }(e_1,e_2,\ldots,e_k)\triangleq\dot\max (e_1,e_2,\ldots,e_k)$
	\end{itemize}
    
\item $\forall e_1,e_2 \in  Expr_{\mathcal{S}}$,
	\begin{itemize}
    \item $f_{\dot\min}(\dot-e_1,\dot-e_2) \triangleq \dot-(f_{\dot\max }(e_1,e_2))$
    \item $f_{\dot\max }(\dot-e_1,\dot-e_2) \triangleq \dot-(f_{\dot\min}(e_1,e_2))$
    \end{itemize}
    
\item $\forall e_1,e_2 \in  Expr_{\mathcal{S}}, \forall c\in\mathbb{R}$, 
	\begin{itemize}
    \item $f_{\dot\min}(c\dot\cdot e_1,c\dot\cdot e_2)\triangleq
       \begin{cases}
         c\dot\cdot f_{\dot\min}(e_1,e_2), \text{ if } c\geq0 \\
         c\dot\cdot f_{\dot\max }(e_1,e_2), \text{ if } c<0
        \end{cases}$
        
     \item $f_{\dot\max }(c\dot\cdot e_1,c\dot\cdot e_2) \triangleq
      \begin{cases}
          c \dot\cdot  f_{\dot\max }(e_1,e_2), \text{ if } c\geq0 \\
          c \dot\cdot  f_{\dot\min}(e_1,e_2), \text { if } c<0
      \end{cases}$
    \end{itemize}
    
\item $\forall e, e_1, e_2 \in Expr_\mathcal{S}$, 
 	\begin{itemize}
 		\item $f_{\dot\min}(e_1\dot\pm e,e_2\dot\pm e) \triangleq f_{\dot\min}(e_1,e_2)\dot\pm e $
 		\item $f_{\dot\max }(e_1\dot\pm e,e_2\dot\pm e) \triangleq f_{\dot\max }(e_1,e_2)\dot\pm e$
 	\end{itemize}
    
\item $\forall e, e_1, e_2 \in Expr_\mathcal{S}$,
	\begin{itemize}
    \item $f_{\dot\min}(e\dot-e_1,e\dot-e_2) = e \dot- f_{\dot\max }(e_1,e_2)$
    \item $f_{\dot\max }(e\dot-e_1,e\dot-e_2) = e \dot- f_{\dot\min}(e_1,e_2)$
    \end{itemize}

\end{enumerate}




\section{A DNA model: bound equations}\label{app:lowerx2}
Below, we give the equation of the lower bound on species $x_2$ from our running DNA model example. According to Def.\ref{def:red}, the derivative of lower bound on the concentration of $x_2$ is computed by selecting the minimum region-dependent (\emph{i.e.}, local) lower bound, out of the 9 possible cases:

\begin{equation*}
\frac{d\underline{x_2}}{dt}= \min(t_\downarrow^{1,1},t_\downarrow^{1,2},t_\downarrow^{1,3},t_\downarrow^{2,1},t_\downarrow^{2,2},t_\downarrow^{2,3},t_\downarrow^{3,1},t_\downarrow^{3,2},t_\downarrow^{3,3})
\end{equation*}
with 


\begin{equation*}
    \begin{split}
    t_\downarrow^{1,1}= &\max (k_1 \underline{x_1}^2,k_{-2} (\frac{DNA_0}{\epsilon} - \frac{k_{-1}}{k_2})) - \\
    &(1+\epsilon) k_{-1} \cdot\min(\underline{x_2}, \frac{M_0-\underline{x_1}}{2},\frac{M_0}{2}-DNA_0-\frac{\underline{x_1}}{2}-\epsilon \frac{k_{-1}}{k_2});
    \end{split}
\end{equation*}


\begin{equation*}
\begin{split}
t_\downarrow^{1,2}= &\max (k_1  \underline{x_1}^2,k_{-2} \frac{DNA_0-\overline{x_3u}}{\epsilon}) - \\
&(1+\epsilon)  k_2 \cdot \min(\underline{x_2},\frac{M_0-\underline{x_1}}{2},\frac{M_0-2  DNA_0-\underline{x_1}+2 \overline{x_3}}{2}) \cdot\\
&\cdot\min(\overline{x_3}, DNA_0) 
\end{split}
\end{equation*}

\begin{equation*}
    \begin{split}
t_\downarrow^{1,3}=& \max (k_1 \underline{x_1}^2, (\frac{k_{-2}}{\epsilon} (DNA_0-\frac{k_{-1}}{\epsilon  k_2}),\frac{k_{-2}}{\epsilon}(DNA_0-\overline{x_3}))  -\\ &\min(\underline{x_2},\frac{M_0-\underline{x_1}}{2}+\min(\overline{x_3}-DNA_0,\frac{k_2}{\epsilon k_{-1}}-DNA_0)))\cdot \\
&(k_2\cdot \min(\overline{x_3},DNA_0)+k_{-1});
\end{split}
\end{equation*}

\begin{equation*}
\begin{split}
t_\downarrow^{2,1}= &k_{-2}\cdot \max (\underline{x_4},DNA_0-\frac{\epsilon k_{-1}}{k_2 }) - \\ &(1+\epsilon)  k_{-1}\cdot \min(\underline{x_2},\frac{M_0}{2}-\max(\underline{x_4},DNA_0-\epsilon \frac{k_{-1}}{k_2}))
\end{split}
\end{equation*}

\begin{equation*}
    \begin{split}
t_\downarrow^{2,2}=&k_{-2}\cdot \max (\underline{x_4},DNA_0-\overline{x_3}) - \\
&(1+\epsilon)  k_2\cdot  \min(\underline{x_2},\frac{M_0}{2}-\underline{x_4},\frac{M_0}{2}-DNA_0+\overline{x_3})\cdot\\ &\min(\overline{x_3},DNA_0-\underline{x_4})
\end{split}
\end{equation*}

\begin{equation*}
    \begin{split}
t_\downarrow^{2,3}= &k_{-2}\cdot \max (\underline{x_4}, DNA_0-\overline{x_3},DNA_0-\frac{k_{-1}}{\epsilon k_2})) -\\
&\min(\underline{x_2},\frac{M_0}{2}-\max(\underline{x_4},DNA_0-\overline{x_3},DNA_0-\frac{k_{-1}}{\epsilon k_2}))\\
       &(k_{-1}+k_2 \cdot \min(\overline{x_3},DNA_0-\underline{x_4}));
\end{split}
\end{equation*}

\begin{equation*}
    \begin{split}
t_\downarrow^{3,1}=&\max (\epsilon k_{-2} \underline{x_4},k_1  \underline{x_1}^2) + \max (\epsilon k_1 \underline{x_1}^2, k_{-2}\cdot (DNA_0-\epsilon k_{-1} \frac{\underline{x_2}}{k_2} )) - \\
       &(1+\epsilon) k_{-1}\cdot \min(\underline{x_2},\frac{M_0-\underline{x_1}}{2}-\epsilon k_1 \cdot {\underline{x_1}^2}{k_{-2}});
\end{split}
\end{equation*}

\begin{equation*}
    \begin{split}
t_\downarrow^{3,2}= &\max (k_1  \underline{x_1}^2,\epsilon k_{-2}\underline{x_4},\epsilon k_{-2}(DNA_0-\overline{x_3})) +k_{-2}\cdot \max (\underline{x_4},DNA_0-\overline{x_3},\epsilon k_1  \frac{\underline{x_1}^2}{k_{-2}}) \\ - 
&(1+\epsilon) k_2 \cdot \min(\underline{x_2},\frac{M_0-\underline{x_1}}{2} -\max(\underline{x_4},\epsilon k_1  \frac{\underline{x_1}^2}{k_{-2}},DNA_0-\overline{x_3})))\cdot \\
&\min(\overline{x_3},DNA_0-\max(\underline{x_4},\epsilon k_1  \frac{\underline{x_1}^2}{k_{-2}}));
\end{split}
\end{equation*}

\begin{equation*}
    \begin{split}
t_\downarrow^{3,3}=&k_1  \underline{x_1}^2 + 
       k_{-2}\cdot \max (\underline{x_4},DNA_0-\overline{x_3})  - \\
       &k_2 \cdot \min(\underline{x_2},\frac{M_0-\underline{x_1}}{2}-\max( \underline{x_4},DNA_0-\overline{x_3}))\cdot \min(\overline{x_3},DNA_0-\underline{x_4}) - \\
       &k_{-1}\cdot \min(\underline{x_2},\frac{M_0-\underline{x_1}-2\cdot \underline{x_4}}{2});
\end{split}
\end{equation*}

\end{appendices}
\end{document}